\documentclass[trackchanges]{aastex7}

\usepackage{amsmath}
\usepackage{CJKutf8}
\usepackage{anyfontsize,graphicx}

\usepackage{CJKutf8}

\begin{document}
\begin{CJK*}{UTF8}{gbsn}

\title{Energy-Containing Electrons in Solar Flares: Improving Hard X-Ray and EUV Diagnostics}

\correspondingauthor{Yingjie Luo}
\email{Yingjie.Luo@glasgow.ac.uk}

\author[0000-0002-5431-545X]{Yingjie Luo (骆英杰)}
\affiliation{School of Physics \& Astronomy,
University of Glasgow
G12 8QQ, Glasgow, UK}
\email{Yingjie.Luo@glasgow.ac.uk} 

\author[0000-0002-8078-0902]{Eduard P. Kontar}
\affiliation{School of Physics \& Astronomy,
University of Glasgow
G12 8QQ, Glasgow, UK}
\email{Eduard.Kontar@glasgow.ac.uk} 

\author[0000-0002-2651-5120]{Debesh Bhattacharjee}
\affiliation{School of Physics \& Astronomy,
University of Glasgow
G12 8QQ, Glasgow, UK}
\email{debusolar624@gmail.com}

\begin{abstract}

Solar flares effectively accelerate particles to non-thermal energies. These accelerated electrons are responsible for energy transport and subsequent emissions in HXR, radio, and UV/EUV radiation. Due to the steeply decreasing electron spectrum, the electron population and consequently the overall flare energetics, are predominantly influenced by low-energy non-thermal electrons. However, deducing the electron distribution in this energy-containing range remains a significant challenge. In this study, we apply the warm-target HXR emission model with kappa-form injected electrons to two well-observed GOES M-class flares. Moreover, we utilize EUV observations to constrain the flaring plasma properties, which enables us to determine the characteristics of accelerated electrons across a range from a few keV to tens of keV. We demonstrate that the warm-target model reliably constrains the properties of flare-associated electrons, even accounting for the uncertainties that had previously been unaddressed. The application of a kappa distribution for the accelerated electrons allows for meaningful comparisons with electron distributions inferred from EUV observations, specifically for energy ranges below the detection threshold of RHESSI. Our results indicate that the accelerated electrons constitute only a small fraction of the total electron population within the flaring region. Moreover, the physical parameters, such as electron escape time and acceleration time scale, inferred from both the warm-target model and the EUV observations further support the scenario in which electrons undergo thermalization within the corona. This study highlights the effectiveness of integrating the warm-target model with EUV observations to accurately characterize energy-containing electrons and their associated acceleration and transport processes.

\end{abstract}


\section{Introduction} 
\label{sec:intro}
Solar flares are the most explosive energy-release events in the solar system. During solar flares, a massive amount of magnetic energy is converted into kinetic energy, leading to the acceleration of electrons to nonthermal energies \citep{1976SoPh...50..153L,2011SSRv..159..107H,2012ApJ...759...71E}, and into thermal energy through plasma heating processes \citep{2010ApJ...725L.161C,2014ApJ...793...70M,2011ApJ...735...42B,2012ApJS..202...11R,2015A&A...584A..89J,2023ApJ...946...53S,2025ApJ...983...58P}. These energy-containing electrons play a crucial role in the processes of energy release, transport, and transformation during solar flares \citep{2002SSRv..101....1A,2017LRSP...14....2B}, and are associated with radiation across multiple wavelengths, including radio, hard X-rays (HXR), and ultraviolet/extreme ultraviolet (UV/EUV) bands \citep{2011SSRv..159...19F,2016A&A...588A.116W}. Considering the nonthermal electron spectrum decreases rapidly---typically with a power-law index greater than 4 \citep{2011SSRv..159..107H,2011SSRv..159..301K,2005A&A...435..743S}---electrons at relatively low energies play a dominant role in determining the total nonthermal energy budget and are critical for assessing flare energy release mechanisms \citep{2012ApJ...759...71E,2013ApJ...764....6O,2023ApJ...947L..13K}. However, constraining the distribution of these energy-containing electrons, particularly at low energies, remains challenging. A key difficulty lies in the fact that electrons at different energies are best detected using different instruments operating at different wavelengths, creating a diagnostic gap. High-energy nonthermal electrons are primarily observed via hard X-ray (HXR) and radio diagnostics. HXR spectroscopy is particularly effective for determining nonthermal electron distributions, as the bremsstrahlung emission is proportional to electron number density and relatively unaffected by transport effects \citep{2011SSRv..159..301K,2011SSRv..159..107H}. Nonetheless, accurately determining the electron distribution in the energy range where thermal emission dominates remains difficult due to instrumental limitations—such as spectral resolution, pile-up effects, and the use of attenuators that affect low-energy response \citep{2011SSRv..159..301K,2002SoPh..210...33S}.
 At lower temperatures, UV/EUV observations, such as those from the Atmospheric Imaging Assembly \citep[AIA,][]{2012SoPh..275...17L} on board the Solar Dynamics Observatory \citep[SDO,][]{2012SoPh..275....3P}, are sensitive to thermal plasma below $\sim$20~MK. However, these observations are insensitive to high-energy nonthermal electrons and provide limited constraints on suprathermal tails \citep{2015ApJ...802...53A}. Moreover, combining multi-wavelength diagnostics introduces additional complexity due to differences in temporal, spatial, and spectral coverage, making it difficult to reconstruct a complete and continuous electron energy spectrum \citep{2016A&A...588A.116W}.

Several studies \citep{2015ApJ...815...73B,2014ApJ...789..116I,2015Ge&Ae..55..995M,2019ApJ...872..204B} have attempted to combine HXR and EUV observations to better constrain electron distributions across a broader energy range. HXR spectral analysis using different emission models, such as isothermal fits, cold-target thick-target bremsstrahlung, and warm-target models, can provide insights into both thermal and nonthermal electron distributions. Meanwhile, multi-channel observations from SDO/AIA enable the reconstruction of the differential emission measure (DEM) as a function of temperature, allowing thermal plasma diagnostics in the range of approximately 0.1 to 20~MK \citep{2012SoPh..275...41B,2015ApJ...802...53A} through various inversion algorithms \citep{2012A&A...539A.146H,2015ApJ...807..143C}. \citet{2015ApJ...815...73B} compared the mean electron flux spectra derived independently from Reuven Ramaty High Energy Solar Spectroscopic Imager \citep[RHESSI;][]{2002SoPh..210....3L} HXR spectra and AIA-based DEMs, and found significant discrepancies—particularly at low energies—highlighting the difficulty in constructing a consistent electron distribution across instruments. Further studies \citep{2015Ge&Ae..55..995M,2014ApJ...789..116I,2019ApJ...872..204B} performed simultaneous forward-fitting on spatially consistent RHESSI and AIA data using instrument response functions. These efforts showed that combined fitting can improve constraints on the electron distribution, primarily for thermal plasma parameters, but still provide limited information about the nonthermal electron population at low energies (below $\sim$20~keV). This limitation arises mainly because the commonly used cold-target model does not reliably constrain nonthermal electrons in flare conditions.

The cold-target model assumes $E \gg k_\text{B}T$, where $k_\text{B}$ is the Boltzmann constant and $T$ is the plasma temperature. However, this condition is not always satisfied in flare loops, where temperatures often exceed $10^7$~K, rendering the cold-target approximation invalid for electrons with energies around 10--20~keV. As a result, the cold-target model fails to accurately describe the physical processes governing electron transport at these energies, making it ineffective for constraining the low-energy portion of the nonthermal electron spectrum \citep{2011SSRv..159..107H}---precisely the region most crucial for flare energetics. To address this limitation, the warm-target model (WTM) was developed \citep{2015ApJ...809...35K,2019ApJ...871..225K}. This model accounts for the thermalization of accelerated electrons in hot, dense coronal plasma and incorporates collisional diffusion under warm-target conditions ($E \sim k_\text{B}T$), offering a more realistic treatment of electron energy loss and transport. \citet{2019ApJ...871..225K} showed that this approach allows accurate determination of the power-law-form electron spectrum, especially the low-energy cutoff of the nonthermal electron distribution which is key to the accurate estimation of electron power. A good understanding of these parameters is also necessary to better characterize the time evolution of flare-accelerated electrons \citep{2025ApJ...987..211B}. Furthermore, \citet{2024ApJ...974..119L} applied a kappa-form electron distribution within the warm-target framework to characterize flare-accelerated electrons. The kappa distribution, often interpreted as the result of stochastic acceleration processes \citep{2014ApJ...796..142B,2021PhRvL.126m5101A}, provides a physically motivated description of flare-accelerated electrons that naturally spans over the thermal-to-nonthermal transition. The utilization of the kappa-form electron distribution in HXR spectral analysis has been demonstrated to be successful \citep{2009A&A...497L..13K,2013ApJ...764....6O,2015ApJ...799..129O,2015ApJ...815...73B,2017ApJ...835..124E,2024ApJ...974..119L}. Once determined by the observed HXR spectrum, the kappa distribution specifies the electron distribution across the full energy range, including below the instrument's observation range (e.g., $\sim$3~keV for RHESSI, depending on attenuator state). In this way, it also naturally covers the range accessible to EUV diagnostics. This makes it particularly well suited for direct comparison with EUV-derived electron distributions, significantly improving the overall constraints on the flare electron population.

By precisely determining the electron distributions of both nonthermal and thermal components during solar flares, we can better analyze the processes of energy release, transformation, and transport \citep{2004JGRA..10910104E,2011SSRv..159..107H}. One key diagnostic is the ratio of nonthermal accelerated electron number density to that of the ambient thermal population. Observational studies indicate that nonthermal electron densities are typically much lower—often of the order of 1\%—even during the impulsive phase of flares \citep{2013ApJ...764....6O,2023ApJ...947L..13K}. In addition, characteristic timescales such as the electron escape time and acceleration time can be inferred from both observational data and theoretical models \citep{2013ApJ...777...33C,2014ApJ...780..176K}. Together, these parameters provide critical insight into how magnetic energy is partitioned and redistributed throughout the flaring coronal environment.

These considerations highlight the need for a comparative diagnostic approach using EUV and HXR observations to better characterize energy-containing electrons and understand their role in flare energetics. In this study, we investigate two well-observed GOES M-class limb flares to determine the properties of energy-containing electrons across a broad energy range. We apply the warm-target model with a kappa-form electron distribution to analyze RHESSI HXR spectra and derive the characteristics of flare-accelerated electrons. We also demonstrate the effectiveness of the warm-target model in reliably determining the electron distribution, even when the thermal loop parameters—such as temperature and density—are constrained only within plausible observational bounds. Meanwhile, we use EUV observations from SDO/AIA to reconstruct the DEM distribution of the thermal electron population. Unlike previous RHESSI-based kappa studies that relied on the cold-target assumption, our use of the warm-target model allows a more reliable determination of the accelerated electron population. We show that the derived ratio suggests that the accelerated electrons constitute only a small fraction. In addition, we derive key timescales related to electron acceleration and transport. This combined analysis enables a more comprehensive understanding of flare energetics and the evolution of flare-associated electrons. The paper is organized as follows. Section~\ref{sec:20110224} presents the analysis of the 2011 February 24 M3.5-class flare. Section~\ref{sec:20120719} describes the corresponding results for the 2012 July 19 M7.7-class flare. In Section~\ref{sec:conclusion}, we summarize our findings.

\section{2011 February 24 M3.5 solar flare} \label{sec:20110224}

\begin{figure*}[!ht]
\plotone{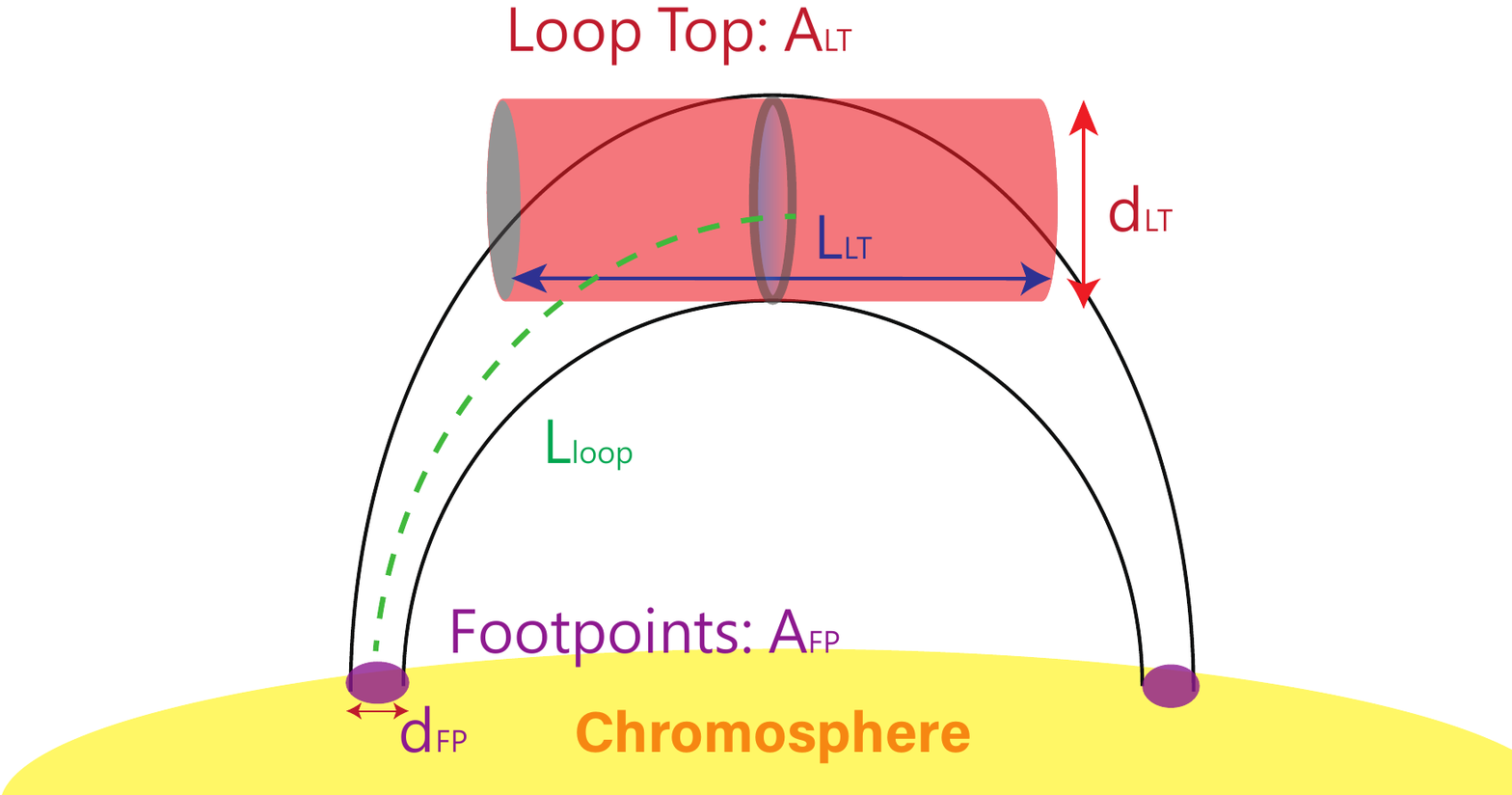}
\caption{Schematic representation of flare coronal geometry, illustrating the key parameters used throughout this study. The diagram shows the loop-top source with diameter $d_{\rm LT}$ and length $L_{\rm LT}$, footpoint diameter $d_{\rm FP}$, and half-loop length $L_{\rm loop}$. The corresponding injection areas ($A_{\rm LT}$ and $A_{\rm FP}$) and volume $V_{\rm LT}$ are used in estimating electron densities and fluxes from both HXR and EUV observations. This schematic serves as a general reference for the geometric configuration assumed in both flare events analyzed.}
\label{fig: sche_fig}
\end{figure*}

The two flares studied in this project exhibit typical loop-top soft X-ray (SXR) sources accompanied by HXR footpoints. We utilize the geometric parameters derived from X-ray and EUV imaging observations for subsequent calculations. A schematic diagram (Figure \ref{fig: sche_fig}) is provided to illustrate the key parameters used throughout the analysis.

\begin{figure*}[!ht]
\plotone{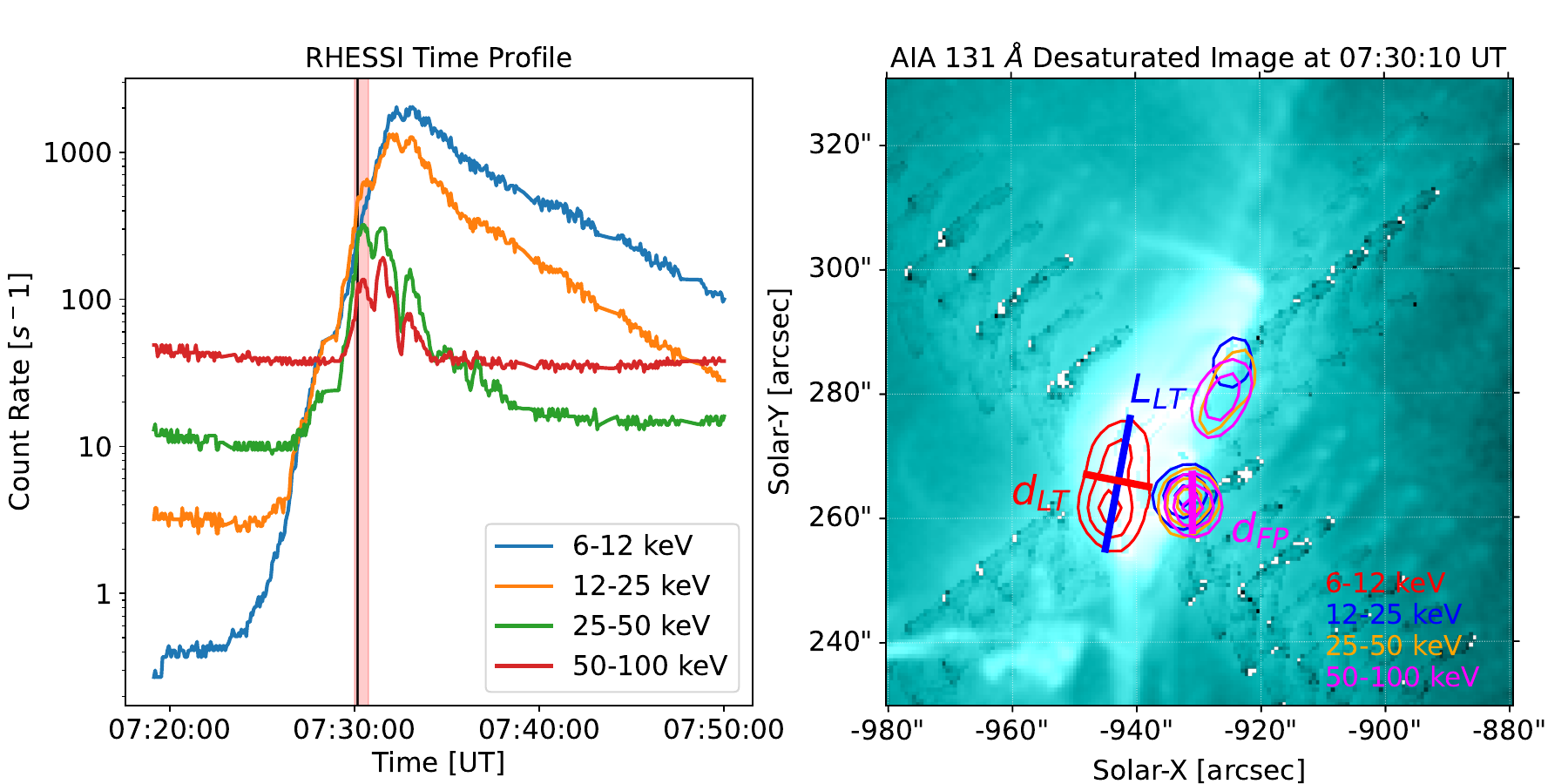}
\caption{Left panel: RHESSI light curves showing the flare evolution. The red shaded region marks the HXR burst interval used for spectral analysis (07:30:00--07:30:44~UT), while the black shaded region indicates the time used for DEM reconstruction (07:30:10~UT). Right panel: AIA 131\,\AA\ image at 07:30:10~UT, de-saturated for clarity, overlaid with RHESSI HXR contours at the 50\%, 70\%, and 90\% levels in four energy bands (6--12, 12--25, 25--50, and 50--100\,keV). The contours highlight the loop-top and footpoint HXR sources used in geometric analysis. The geometric parameters $d_{\rm{LT}}$, $L_{\rm{LT}}$, and $d_{\rm{FP}}$ are marked as red, blue, and magenta lines, respectively, and are used for volume and area estimates.}
\label{fig: event_0224}
\end{figure*}

The first event analyzed is a GOES M3.5-class flare. The HXR burst under investigation occurred between 07:30:00~UT and 07:30:44~UT (red shaded region in Figure~\ref{fig: event_0224}, left panel), preceding the GOES SXR peak at approximately 07:35~UT. The flare was located near the eastern limb of the solar disk (Figure~\ref{fig: event_0224}, right panel). HXR spectroscopic imaging reveals a characteristic loop-top SXR source accompanied by two HXR footpoint sources (Figure~\ref{fig: event_0224}, right panel). Following the schematic in Figure~\ref{fig: sche_fig}, we summarize the geometric parameters for the 2011 February 24 flare, as derived from HXR imaging and AIA observations. Based on 50\% CLEAN source contours, the related source size parameters are summarized in Table \ref{tab: obs_params_20110224}.

\begin{table}
    \centering
    \renewcommand{\arraystretch}{1.3}
    \begin{tabular}{||c|c|p{6.5cm}||}
        \hline
        \textbf{Parameter} & \textbf{Value} & \textbf{Remark} \\
        \hline\hline
        \multicolumn{3}{||c||}{\textit{Observation size}} \\
        \hline
        Looptop source diameter $d_{\rm LT}$ 
        & $10.2\arcsec \ (= 7.3~\mathrm{Mm})$ 
        & Defines column depth for AIA-DEM. \\
        Looptop source length $L_{\rm LT}$ 
        & $21.4\arcsec \ (= 15.4~\mathrm{Mm})$ 
        & Used to compute looptop source volume. \\ 
        \shortstack[c]{Looptop cross-sectional area\\ $A_{\rm LT} = \pi (d_{\rm LT}/2)^2$}
        & $4.2 \times 10^{17}$ cm$^2$ 
        & Defines electron injection area at the looptop. \\
        \shortstack[c]{Looptop source volume\\ $V_{\rm LT} = \pi (d_{\rm LT}/2)^2 L_{\rm LT}$}
        & $6.5 \times 10^{26}$ cm$^3$ 
        & Used for density and emission measure estimates. \\
        Footpoint source diameter $d_{\rm FP}$ 
        & $9.0\arcsec \ (= 6.5~\mathrm{Mm})$ 
        & Defines HXR footpoint source size. \\
        \shortstack[c]{Footpoint cross-sectional area\\ $A_{\rm FP} = \pi (d_{\rm FP}/2)^2$}
        & $3.3 \times 10^{17}$ cm$^2$ 
        & Defines electron injection area at the footpoint. \\
        Half-loop length $L_{\rm loop}$ 
        & $15.8$ Mm 
        & Geometric parameter for WTM fit. \\
        \hline

        \multicolumn{3}{||c||}{\textit{AIA-DEM-derived thermal properties}} \\
        \hline
        Emission measure per area $EM_A$ 
        & $6.96 \times 10^{29}$ cm$^{-5}$ 
        & Used to compute electron number density.\\
        Average temperature $T_{\rm avg}$ 
        & $17.9$ MK ($1.54$ keV) 
        & Input for WTM loop thermal parameters. \\
        \shortstack[c]{Electron number density\\ $n_e = \sqrt{EM_A / d}$ \; (with $d=d_{\rm LT}$)} 
        & $3.1 \times 10^{10}$ cm$^{-3}$ 
        & Input for WTM loop thermal parameters. \\
        Plausible temperature range 
        & $1.31$--$1.61$ keV (15.2--18.7 MK) 
        & Used to assess uncertainty of WTM spectral fits. \\
        Plausible density range 
        & $(2.6$--$4.7)\times 10^{10}$ cm$^{-3}$ 
        & Used to assess uncertainty of WTM spectral fits. \\
        \hline

        \multicolumn{3}{||c||}{\textit{HXR-derived thermal properties}} \\
        \hline
        Isothermal component temperature
        & $1.30$ keV (15.1 MK) 
        & Used in WTM fit. \\
        Isothermal component electron density 
        & $4.4 \times 10^{10}$ cm$^{-3}$ 
        & Used in WTM fit. \\
        Plausible density range
        & ($3.6$--$5.1) \times 10^{10}$ cm$^{-3}$ 
        & Used to assess uncertainty of WTM spectral fits. \\
        Plausible temperature range 
        & $1.16$--$1.44$ keV (13.5--16.7 MK) 
        & Used to assess uncertainty of WTM spectral fits. \\
        \hline
    \end{tabular}
    \caption{Observation size, AIA-DEM-derived, and HXR-derived thermal parameters for the 2011 February 24 flare. These values provide the geometric and thermal constraints used in the warm-target model analysis, with remarks indicating their specific roles.}
    \label{tab: obs_params_20110224}
\end{table}

\subsection{DEM Analysis} \label{sec:20110224 dem}

\begin{figure*}[!ht]
\plotone{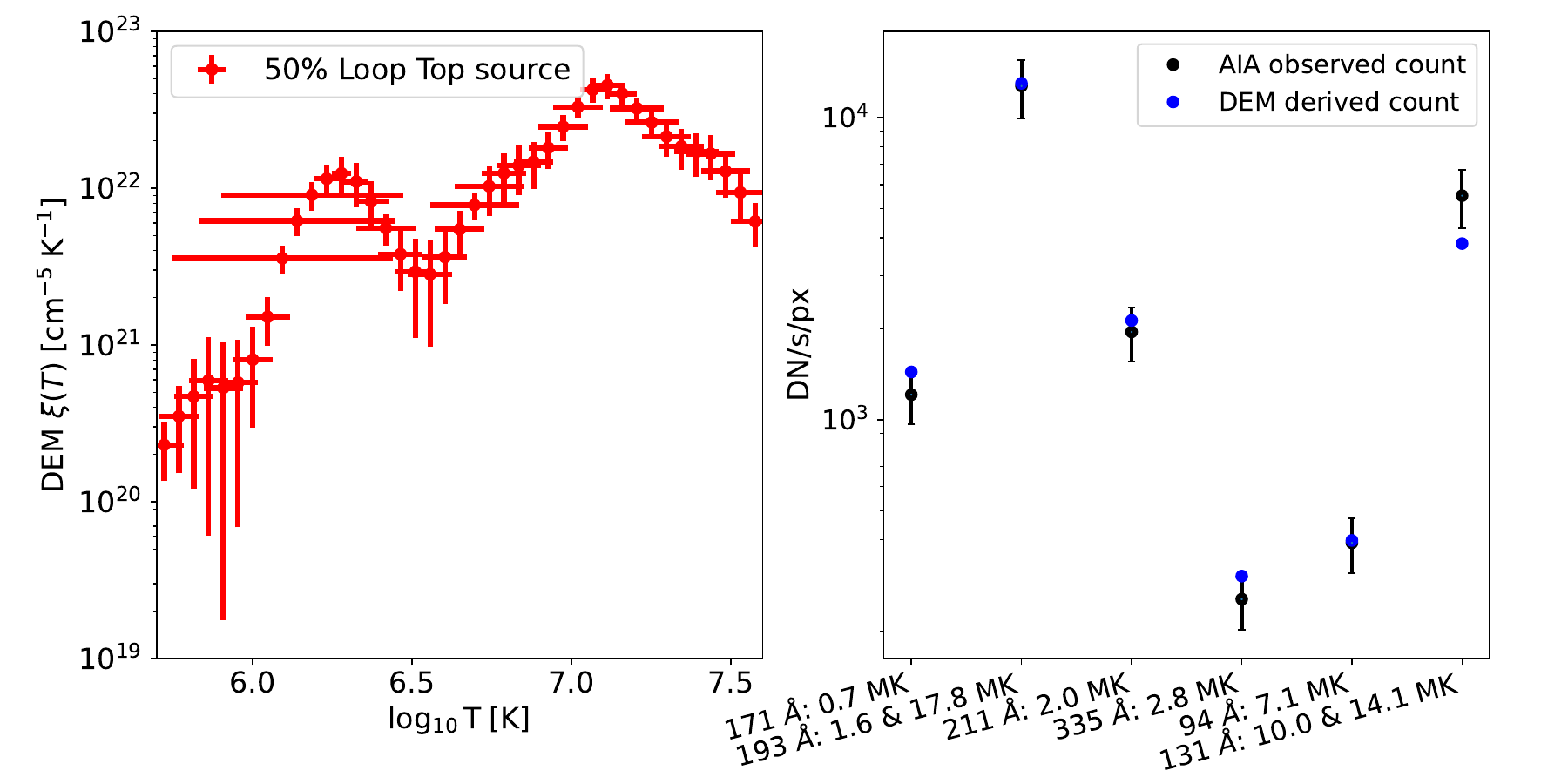}
\caption{DEM analysis of the 50\% loop-top source region for the 2011 February 24 flare using AIA EUV observations. Left panel: Reconstructed DEM distribution $\xi(T)$ obtained from regularized inversion applied to de-saturated AIA images. The DEM curve shows two peaks: a cooler background component and a hotter flare-heated component. Right panel: Comparison between observed AIA data counts and simulated counts generated from the DEM solution, showing good agreement across the six EUV channels.}
\label{fig: dem_spec_0224}
\end{figure*}

\begin{figure*}[!ht]
\plotone{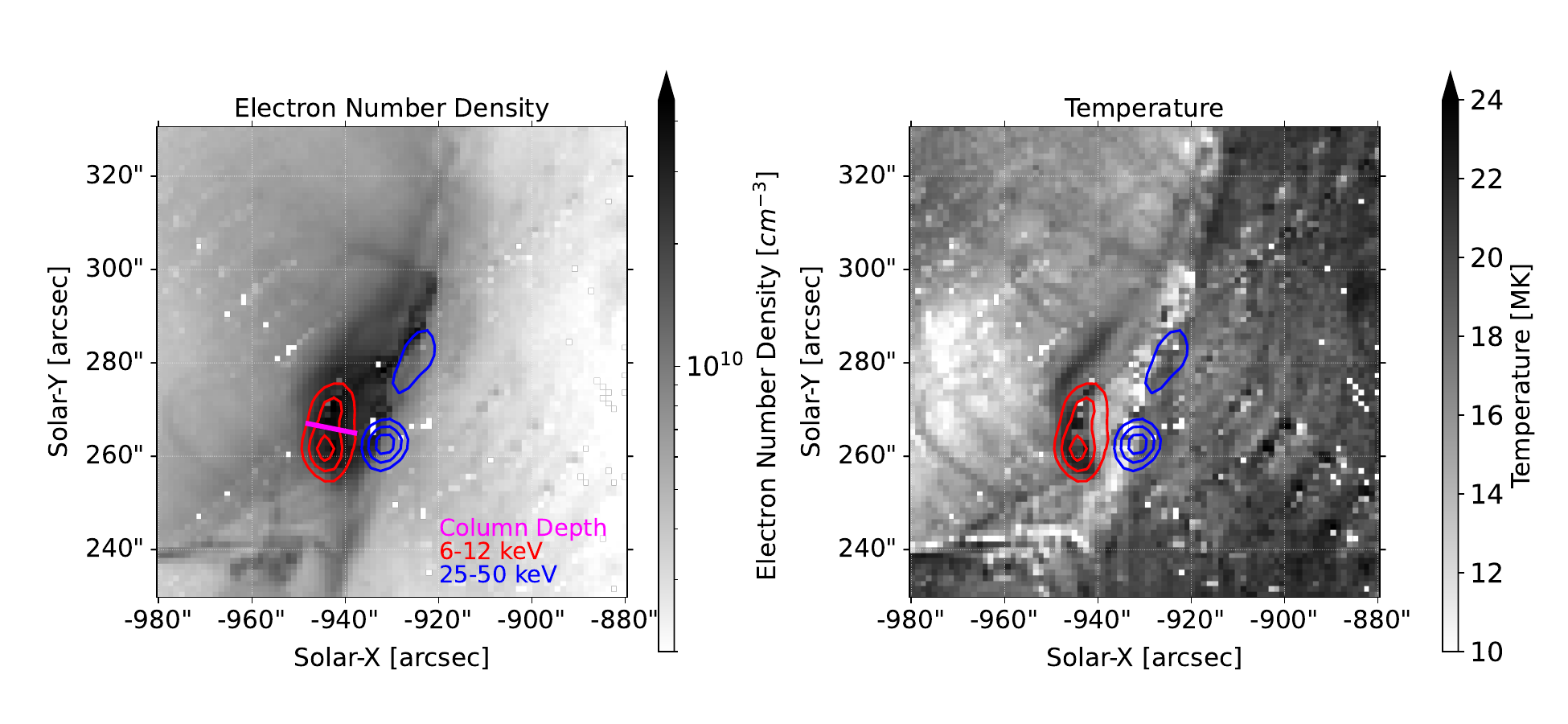}
\caption{Electron number density and temperature maps derived from per-pixel DEM inversion of AIA EUV data for the 2011 February 24 flare. Left panel: Electron number density map, calculated from the total emission measure using a column depth of $d = d_{\rm LT} = 10.2\arcsec = 7.3$~Mm (magenta line). Right panel: DEM-weighted average temperature map.}
\label{fig: dem_map_0224}
\end{figure*}

To analyze the plasma properties at the flare site, we reconstruct the differential emission measure (DEM) distribution as a function of temperature using SDO/AIA observations from six EUV filters (94, 131, 171, 193, 211, and 335~\AA) during the HXR burst ($\sim$07:30:10~UT; black shaded region in Figure~\ref{fig: event_0224}, left panel). During this interval, the 131~\AA\ and 193~\AA\ channels are saturated at the loop-top region due to their high sensitivity to hot plasma. To recover the real data counts in the saturated regions, we apply a de-saturation algorithm from the \texttt{DESAT} package in \textit{SSWIDL} to these two channels. This method uses correlation and diffraction fringe inversion to restore saturated pixels \citep{2014ApJ...793L..23S}. A de-saturated 131~\AA\ image is shown in Figure~\ref{fig: event_0224} (right panel).

We apply the regularized inversion method \citep{2012A&A...539A.146H} to the de-saturated images to derive the DEM per area, $\xi_\text{A}(T) = n^2 \, \frac{dl}{dT}$ [cm$^{-5}$\,K$^{-1}$], as a function of temperature $T$. We begin by analyzing the DEM distribution within the 50\% contour of the loop-top SXR source, which corresponds to the likely site of accelerated electron injection. A systematic uncertainty of 20\% is included, using $\text{DN}_\text{err} = \sqrt{\text{DN} + (0.2\,\text{DN})^2}$ to account for instrumental and calibration errors. In addition, for the 131~\AA\ and 193~\AA\ channels where de-saturation was applied, an extra 10\% uncertainty was added to the data counts to reflect the possible error from reconstruction process. For this event, we choose to run the regularization once, working with the constraint matrix derived from the minimum of the ``EM loci'' \citep{2007ApJ...655..598C,2007ApJ...658L.119S}. This approach prevents the potential underestimation of the total emission measure caused by excessive smoothing of $\xi(T)$, as discussed by \citet{2023A&A...672A.120M}. The reconstructed DEM distribution and the corresponding AIA observed versus simulated counts are presented in Figure~\ref{fig: dem_spec_0224}. The simulated counts show good agreement with the observed data, indicating a robust DEM solution (Figure~\ref{fig: dem_spec_0224}, right panel).

The recovered DEM distribution exhibits two distinct peaks: a lower-temperature component, likely arising from background emission along the line of sight \citep{2012ApJ...760..142B}, and a higher-temperature component associated with flare-heated plasma, which is the primary focus of this study. From the DEM curve, we derive a total emission measure of $\text{EM}_\text{A} = \sum_{T_\text{min}}^{T_\text{max}} \xi_\text{A}(T)\,dT = 6.96 \times 10^{29}$~cm$^{-5}$, an average temperature of $T_\text{avg}=(\sum_{T_\text{min}}^{T_\text{max}} T\xi_\text{A}(T) dT)/(\sum_{T_\text{min}}^{T_\text{max}} \xi_\text{A}(T) dT)=$17.9~MK ($k_{B}T_{\rm avg}$=1.54~keV), and an electron number density of $n_e = \sqrt{\text{EM}_\text{A} / d} = 3.1 \times 10^{10}$~cm$^{-3}$, assuming a column depth of $d = d_{\rm LT} = 10.2\arcsec = 7.3$~Mm (magenta line in Figure~\ref{fig: dem_map_0224}, left panel). In addition to the loop-top region, we reconstruct the DEM distribution on a per-pixel (re-binned to 1.2\arcsec) basis across the flare site. The resulting electron number density and temperature maps are shown in Figure~\ref{fig: dem_map_0224}.

It is important to note that while DEM distributions obtained from AIA-only inversion algorithms are frequently used to derive thermal plasma properties, such properties cannot be unambiguously determined because of the limited number of AIA filters, their specific sensitivity ranges, and the fact that several high-temperature channels (e.g., 131~\AA\ and 193~\AA) are also sensitive to cooler plasma components. Previous studies have shown that the high-temperature component inferred from AIA observations may be overestimated, owing to the multi-thermal response of the AIA filters, a conclusion supported by comparisons with other instruments \citep[e.g.,][]{2024ApJ...961..181A,2015ApJ...815...73B}. We account for the associated uncertainties in the analysis that follows.

\subsection{HXR spectral analysis} \label{sec:20110224 hxr}

\begin{table}
    \centering
    \begin{tabular}{||c|c|c|c||}
        \hline
        \textbf{Parameter} 
        & \shortstack[c]{\rule{0pt}{2.6ex}\textbf{$f_{\rm vth}$+WTM}\\\textbf{(thermal properties derived from HXR)}} 
        & \shortstack[c]{\rule{0pt}{2.6ex}\textbf{$f_{\rm vth}$+WTM}\\\textbf{(thermal properties derived from DEM)}} 
        & \textbf{Status} \\
        \hline\hline
        \multicolumn{4}{||c||}{\textit{\boldmath $f_{\rm vth}$ (isothermal component, fixed)}} \\
        \hline
        EM [$10^{49}\,\mathrm{cm}^{-3}$]   & 0.127  & 0.127  & Fixed \\
        $k_{B}T$ [keV]                     & 1.30   & 1.30   & Fixed \\
        \hline
        \multicolumn{4}{||c||}{\textit{WTM component: loop thermal parameters (fixed)}} \\
        \hline
        $n_{\rm loop}$ [$10^{10}\,\mathrm{cm}^{-3}$] & 4.4  & 3.1  & Fixed \\
        $k_{B}T_{\rm loop}$ [keV]                    & 1.30 & 1.54 & Fixed \\
        $L$ [Mm]                                     & 15.8 & 15.8 & Fixed \\
        \hline
        \multicolumn{4}{||c||}{\textit{WTM component: kappa parameters related to accelerated/injected electrons (free)}} \\
        \hline
        $\dot{N}_0$ [$10^{35}\,\mathrm{e^-}\,\mathrm{s}^{-1}$] & $22.5 \pm 0.8$ & $26.1 \pm 0.9$ & Free \\
        $\kappa$                                                & $5.14 \pm 0.04$ & $5.04 \pm 0.04$ & Free \\
        $k_{B}T_{\kappa}$ [keV]                                 & $1.29 \pm 0.04$ & $1.14 \pm 0.04$ & Free \\
        \hline
        \multicolumn{4}{||c||}{\textit{Fit diagnostics}} \\
        \hline
        $\chi^2$      & 2.11 & 2.37 & N/A \\
        Power [$10^{28}\,\mathrm{erg\,s}^{-1}$] & 2.23 & 2.35 & N/A \\
        \hline
    \end{tabular}
    \caption{Spectral fitting results for the 2011 February 24 M3.5-class flare using the warm-target model with a kappa-form injected electron distribution and an isothermal component ($f_{\rm vth}$). Two fits are shown that differ in the thermal properties specified for the WTM: one set derived from an HXR isothermal fit, and the other from DEM analysis of the loop-top region. In both fits, the $f_{\rm vth}$ component adopts the same numerical values and is fixed. The WTM loop thermal parameters are fixed to the corresponding values from the chosen approach, whereas the kappa-distribution parameters ($\dot{N}_0$, $\kappa$, $k_{\rm B}T_\kappa$) are free. Fixed parameters, free parameters, and diagnostic quantities are indicated.}
    \label{tab: 20110224 fit}
\end{table}

\begin{figure*}[!ht]
\plotone{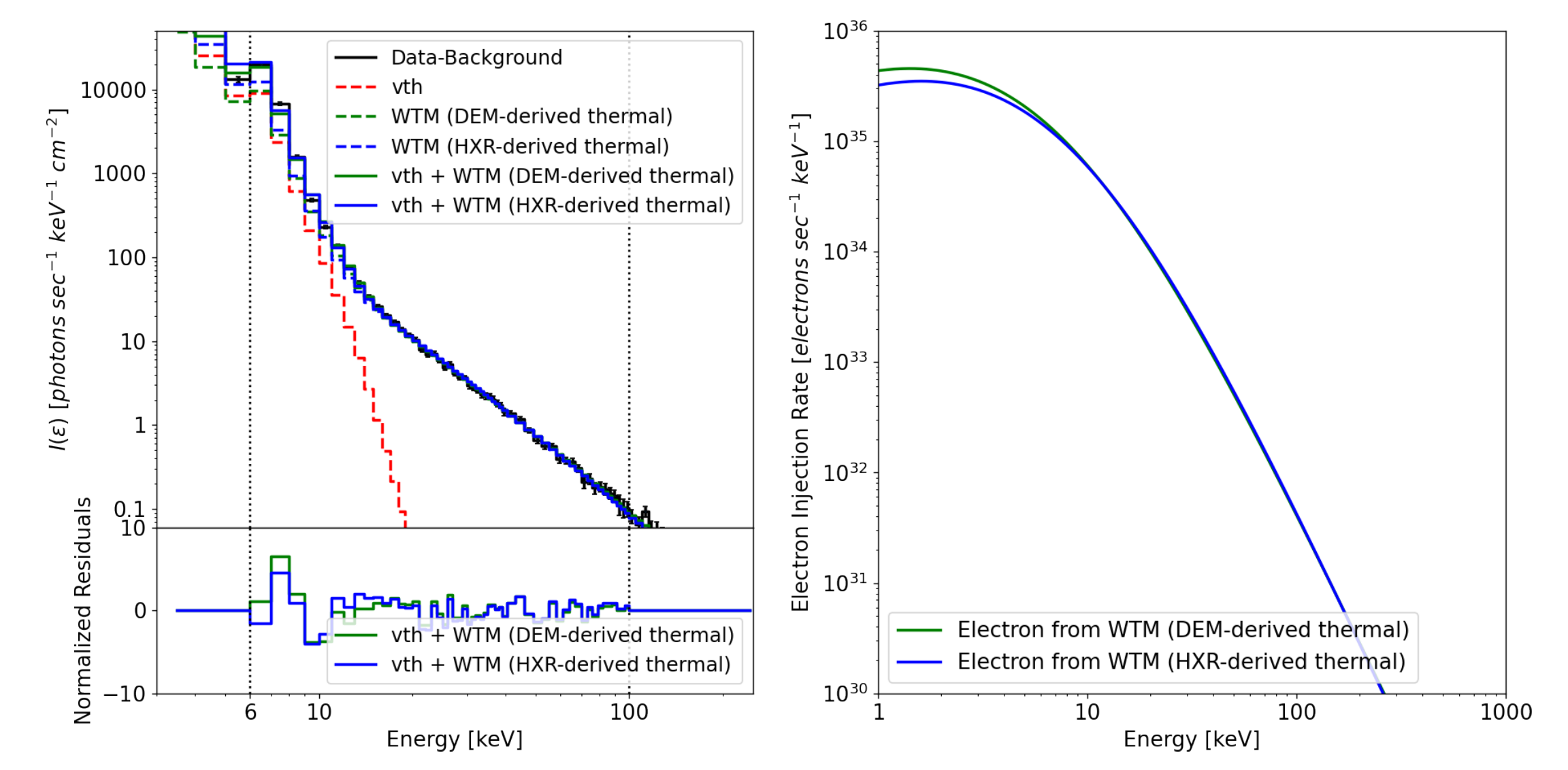}
\caption{HXR spectral fitting results for the 2011 February 24 flare using the warm-target model with a kappa-distributed electron population. Left panel: Fitted photon spectra using thermal input parameters from the isothermal pre-burst fit (blue curves) and from DEM analysis (green curves). Both fits use the same $f_{\rm vth}$ + $f_{\rm thick\_warm\_kappa}$ framework. Normalized residuals for each fit are shown in the bottom panel. Right panel: Corresponding injected electron spectra for the two fits, demonstrating close agreement above 10~keV.}
\label{fig: wtm_res_0224}
\end{figure*}

A detailed HXR spectral analysis using the warm-target model (WTM) for this event is provided in \citet{2024ApJ...974..119L}. As described in that study, applying the WTM requires the thermal properties of the target coronal loop (the half-loop length $L$, number density $n_{\text{loop}}$, and temperature $T_{\text{loop}}$), which determine the thermalization of the injected electrons. In \citet{2024ApJ...974..119L}, these parameters were derived from an isothermal component ($f_{\rm vth}$) fitted to the pre-burst spectrum, yielding $n_{\text{loop}} = 4.4 \times 10^{10}$~cm$^{-3}$, $k_{\rm B} T_{\text{loop}} = 1.30$~keV, and $L = 15.8$~Mm. In this study, we apply the same WTM framework with kappa-form injected electrons, but instead adopt the thermal properties derived from DEM analysis of the loop-top region for comparison. We retain the same half-loop length and use $n_{\text{loop}} = 3.1 \times 10^{10}$~cm$^{-3}$ and $k_{\rm B} T_{\text{loop}} = 1.54$~keV, the electron number density and the averaged temperature based on the DEM results described in Section~\ref{sec:20110224 dem}. We keep the isothermal component fixed from the pre-burst time, focusing on the contributions from the injected electrons within the WTM framework. The result fit parameters are summarized in Table~\ref{tab: 20110224 fit}, including the total electron injection rate (\( \dot{N}_0 \)), the kappa temperature (\( T_\kappa \)), and the kappa index (\( \kappa \)), also the calculated total power carried by nonthermal electrons. The modeled photon spectrum along with the injected electron spectrum is shown in Figure~\ref{fig: wtm_res_0224}. We find that the two sets of fits produce comparable results and electron spectra, particularly in the energy range above 10~keV.

\begin{figure*}[!ht]
\plotone{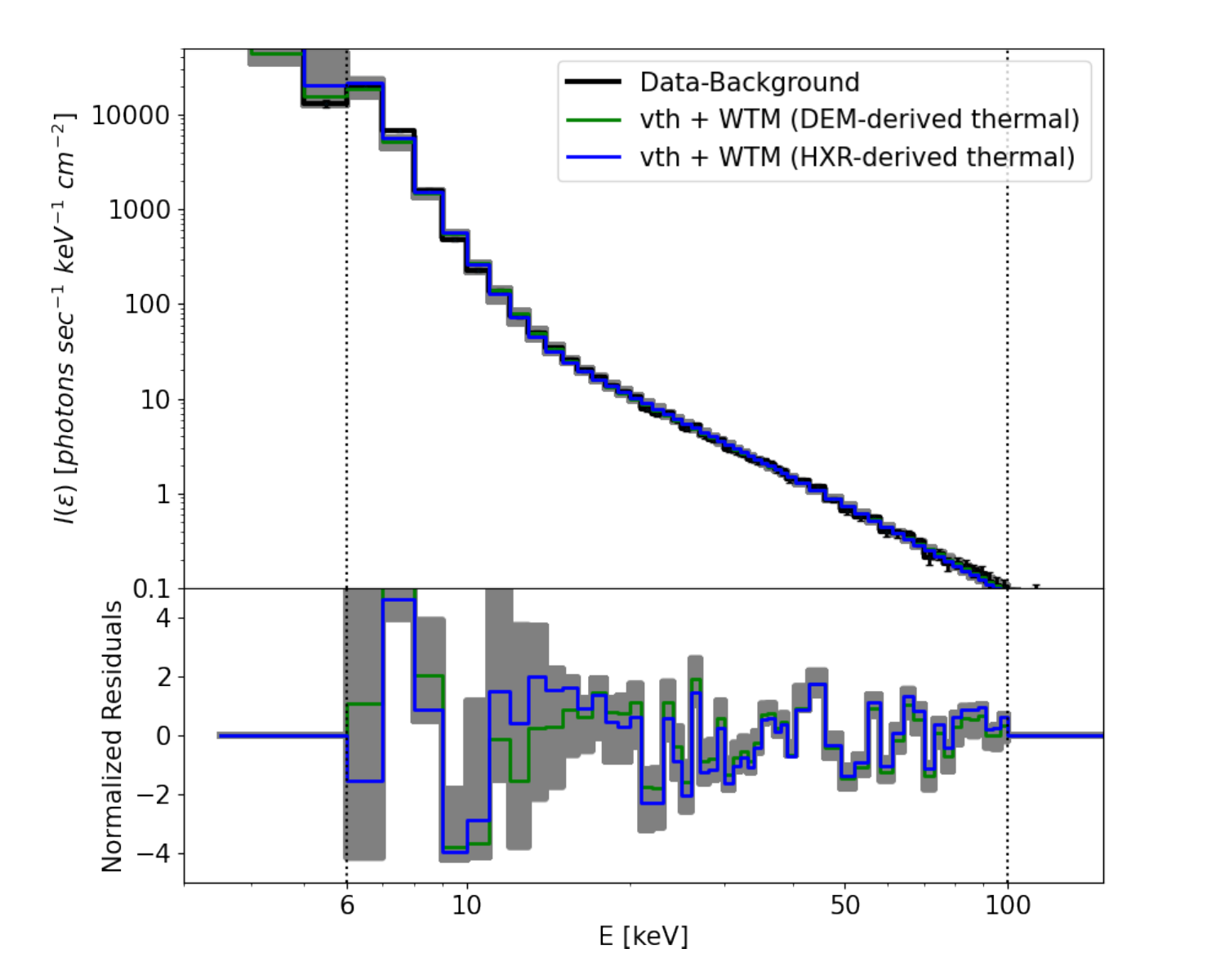}
\caption{HXR photon spectra fitted with the warm-target model using a range of thermal input parameters for the 2011 February 24 flare. The parameters include loop-top temperatures $k_{\rm B}T_{\rm loop} = 1.16$--$1.61$~keV and electron densities $n_{\rm loop} = (2.6$--$5.1) \times 10^{10}$~cm$^{-3}$. The top panel shows the fitted spectra; the bottom panel displays the corresponding normalized residuals. Despite the variation in input parameters, the warm-target model yields consistent and robust fits to the observed RHESSI spectra.}
\label{fig: group_res_0224}
\end{figure*}

We further quantitatively investigate how variations in the thermal input parameters affect the WTM spectral fitting results. We first consider the 1$\sigma$ uncertainties derived from the pre-burst isothermal fit component ($f_{\rm vth}$), performed during the interval 07:29:52–07:30:00 UT (details can be seen in \citealt{2024ApJ...974..119L}). The resulting 1$\sigma$ confidence temperature and emission measure ranges—1.16–1.44 keV and $(0.083$–$0.171)\times10^{49}$ cm$^{-3}$, which corresponds to an inferred thermal electron density range of $(3.6$–$5.1)\times10^{10}$ cm$^{-3}$, respectively. For thermal parameters inferred from EUV observations, as discussed in Section \ref{sec:20110224 dem}，they cannot be precisely determined from AIA data alone. To represent a plausible range of thermal conditions, we use 07:29~UT (one minute prior to the HXR burst) as a lower bound and 07:32~UT (near the SXR peak) as an upper bound for the purposes of our analysis. The DEM results at these times yield a temperature range of 1.31--1.61~keV and an electron density range of $(2.6$--$4.7) \times 10^{10}$~cm$^{-3}$ for the loop-top region. Combining the thermal parameter ranges derived from AIA and HXR, we constrain a representative range of $k_{\rm B} T_{\text{loop}} = 1.16$--1.61~keV and $n_{\text{loop}} = (2.6$--$5.1) \times 10^{10}$~cm$^{-3}$. Accordingly, the impact of potential AIA-only DEM overestimation on the thermal parameters is mitigated through the inclusion of the HXR-derived isothermal component, which effectively constrains the very hot plasma. This range incorporates the uncertainties from both AIA and HXR diagnostics, minimizing the effect of these uncertainties on the derived parameters and providing a more reliable constraint.

Within this range, the mean values of the fitted kappa-distribution parameters are $\dot{N}_0 = 25.9 \times 10^{35}$~electrons~s$^{-1}$, $k_{\rm B} T_\kappa = 1.27$~keV, and $\kappa = 5.11$, with corresponding standard deviations of $12.5 \times 10^{35}$~electrons~s$^{-1}$, 0.25~keV, and 0.06, respectively. While $\dot{N}_0$ appears less well constrained, the relatively large uncertainty is influenced by a few extreme high values arising when both $T$ and $n_{\text{loop}}$ are at their lower limits. We conclude that the warm-target model remains effective in constraining the nonthermal electron distribution, even when the thermal input parameters are varied within observationally plausible ranges derived from both HXR and EUV diagnostics. The corresponding fitted photon spectra and normalized residuals are shown in Figure~\ref{fig: group_res_0224}.

\subsection{Comparison of Electron Spectra from WTM and EUV Diagnostics} \label{sec:20110224 compare}

\begin{figure*}[!ht]
\plotone{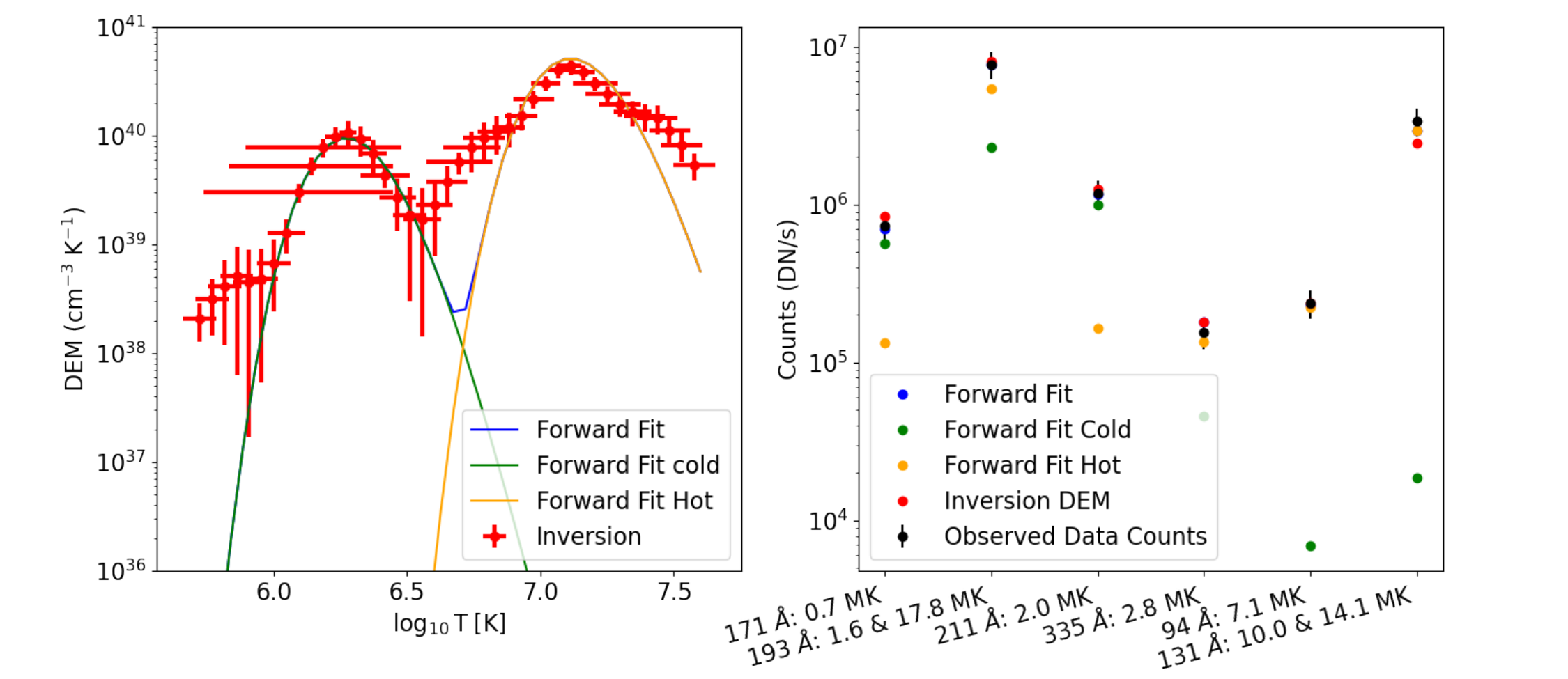}
\caption{Forward-fit results of the DEM distribution for the 2011 February 24 flare loop top region. 
Left panel: DEM distribution derived using two kappa-form components—representing background (cold, green) and flare-heated plasma (hot, orange). The total forward-fitted DEM is shown in blue. For comparison, the DEM obtained from regularized inversion is shown in red. 
Right panel: Comparison between observed AIA data counts (black) and modeled counts from the forward-fit DEM (blue), showing good agreement across all six EUV channels.}
\label{fig: forwardfit_0224}
\end{figure*}

\begin{figure*}[!ht]
\plotone{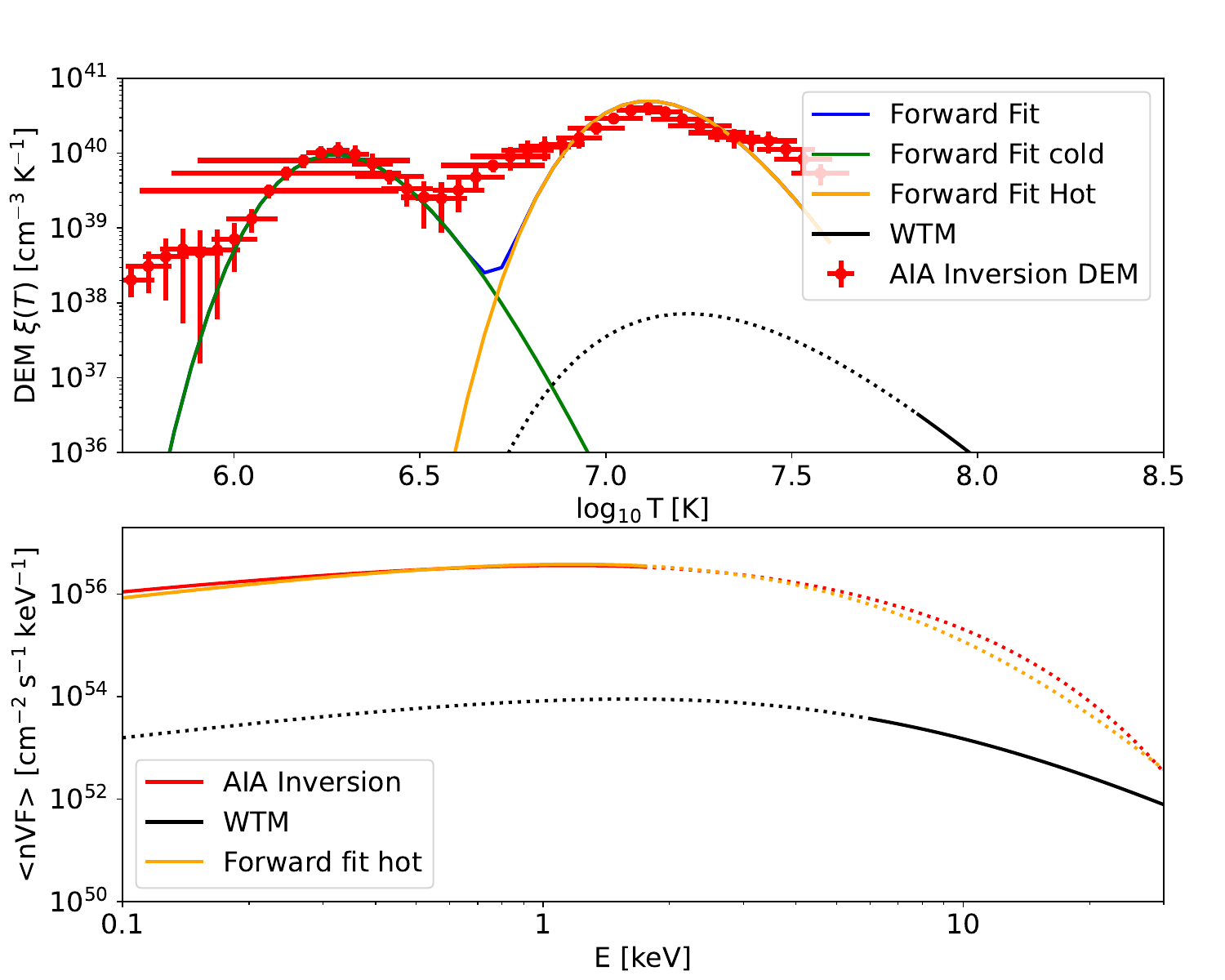}
\caption{Comparison of electron spectra for the 2011 February 24 flare.
Top panel: DEM distributions corresponding to the AIA-derived regularized inversion (red), the forward-fitted hot kappa component (orange), and the WTM-inferred kappa distribution (black). 
Bottom panel: Corresponding mean electron flux spectra $\langle nVF \rangle(E)$ calculated using Equation~\ref{equ: nVF-ra}. Dashed portions of the curves denote regions outside the sensitive ranges of AIA (above $\sim$20~MK) and RHESSI (below 6~keV), where observational constraints are limited.}
\label{fig: spec_comb_0224}
\end{figure*}

In the previous sections, we derived the DEM distribution of the loop-top plasma within AIA's sensitive range, and the kappa-form accelerated electron spectrum across the entire energy range using the WTM. Based on the thick-target model, the derived kappa-form electrons represent the accelerated electrons injected into the flare loops. The DEM distribution of the loop-top region, although it may include some additional material such as from chromospheric evaporation, still provides a reasonable estimate of the total plasma population associated with the acceleration. The relationship between these two distributions offers valuable insight into the flare particle acceleration process. Here, we compare the electron distributions in terms of both the DEM distribution and the mean electron flux spectrum $\langle nVF \rangle(E)$.

We begin by recalling the relationship between the mean electron flux spectrum $\langle nVF \rangle(E)$ and the DEM distribution $\xi(T)$, where $\xi(T) = n^2 \frac{dV}{dT}$ [cm$^{-3}$\,K$^{-1}$] \citep{1988ApJ...331..554B,2013ApJ...779..107B}:

\begin{equation}\label{equ: nVF-ra}
\langle nVF \rangle(E)=\frac{2^{3/2}E}{\sqrt{\pi m_e}} \int_ {0}^{\infty} \frac{\xi(T)}{(k_B T)^{1.5}} \exp(-E/k_B T) \,dT
\end{equation}

To apply this relation to the AIA-derived DEM, $\xi_A(T)$, we multiply by a projection area $A_{\rm proj}$, taken from the 50\% looptop source region, yielding $\xi(T)$ for use in Equation~\ref{equ: nVF-ra}. As discussed in Section~\ref{sec:20110224 dem}, the DEM from regularized inversion reveals two distinct components: a cooler component, likely due to background emission along the line of sight, and a hotter component associated with flare-heated plasma. To focus on the flare-associated plasma, we perform a forward fit of the observed AIA counts using two kappa-form DEM components: a ``cold'' component representing background and a ``hot'' component representing the flare-heated plasma. Each kappa DEM component is defined by three parameters: the total emission measure (EM), the kappa temperature $T_{\kappa}$, and the kappa index $\kappa$, as described by Equation~\ref{equ: kappa-dem2}. Due to the limited number of data points (six AIA channels), we allow all three parameters of the hot component to vary freely, while fixing the kappa temperature of the cold component at 1.8~MK, corresponding to the first peak in the regularized inversion DEM. The emission measure and kappa index of the cold component are kept free. This choice reduces the number of free parameters in the fit and allows us to better constrain the hot component, which is the main focus of the subsequent comparison.

The fit results are as follows: for the cold component, $\text{EM}_\text{cold} = (1.4 \pm 0.3) \times 10^{46}$~cm$^{-3}$, $k_\text{B}T_{\kappa,\rm cold} = 0.16$~keV  (fixed), and $\kappa_\text{cold} = 12.4 \pm 7.2$; for the hot component, $\text{EM}_\text{hot} = (5.9 \pm 1.8) \times 10^{47}$~cm$^{-3}$, $k_\text{B}T_{\kappa,\rm hot} = 1.06 \pm 0.09$~keV, and $\kappa_\text{hot} = 10.2 \pm 7.8$. Most parameters are reasonably well constrained, except for the kappa indices. The less constrained kappa index is expected, as the AIA temperature response functions decline steeply outside their optimal sensitivity ranges, limiting the ability to constrain the high-energy tail of the DEM. The forward-fitted DEM is shown in Figure~\ref{fig: forwardfit_0224} (left), with the total DEM in blue and the cold and hot kappa components in green and orange, respectively. The regularized inversion result is overplotted in red, showing good agreement. The corresponding modeled AIA counts are shown in the right panel and demonstrate strong consistency with the observed data.

The inferred mean electron flux spectra are shown in Figure~\ref{fig: spec_comb_0224} (lower panel). The AIA-derived spectra are computed using Equation~\ref{equ: nVF-ra}, using both the regularized inversion (red curve) and forward-fitted hot component (orange curve). Although the formal integration can be performed across all energies, regions corresponding to temperatures beyond AIA’s sensitivity (above $\sim$20~MK) are shown as dashed lines to indicate increased uncertainty. As discussed in Section \ref{sec:20110224 dem}, the region depicted with dashed curves can hardly be constrained by AIA observations alone. \citet{2013ApJ...779..107B} suggested that the thermal properties above AIA's sensitive range could be significantly higher, as indicated by RHESSI observations. Our analysis of the AIA-derived DEM also suggests a potential underestimation: Table \ref{tab: 20110224 fit} shows that the electron number density inferred from AIA observations is lower than that derived from the RHESSI isothermal fit (3.1 vs.\ 4.4 $\times$ 10$^{10}$ cm$^{-3}$, corresponding to an emission measure ratio of (3.1/4.4)$^{2}\sim0.50$), but the difference is not substantial. We therefore propose that the total emission measured from the hot kappa component can serve as a reasonable lower limit for the thermal plasma present at the flare site.

To obtain a consistent comparison with the AIA-derived DEM, we compute the corresponding DEM distribution and mean electron flux spectrum associated with the kappa-form electron distribution inferred from the WTM. Following \citet{2015ApJ...815...73B}, we use the DEM formulation of kappa-form electron distribution as:
\begin{equation}\label{equ: kappa-dem}
\xi(T) \propto T^{-(\kappa-0.5)} \exp\left(-\frac{\kappa T_\kappa}{T}\right)
\end{equation}

The total emission measure is given by:

\begin{equation}\label{equ: kappa-em}
EM = \int_{0}^{\infty} \xi(T) \,dT \propto \Gamma(\kappa-1.5)(\kappa T_\kappa)^{-(\kappa-1.5)}
\end{equation}

This leads to the explicit expression:

\begin{equation}\label{equ: kappa-dem2}
\xi(T) = \frac{EM \,\kappa^{(\kappa-1.5)} }{\Gamma(\kappa-1.5)T_\kappa} \left(\frac{T_\kappa}{T}\right)^{(\kappa-0.5)} \exp\left(-\frac{\kappa T_\kappa}{T}\right)
\end{equation}

which corresponds to the kappa velocity distribution:

\begin{equation}\label{equ: kappa-dis-v2}
f_k(v)=\frac{n_k}{\pi^{3/2}v_{te}^3\kappa^{3/2}}\frac{\Gamma(\kappa)}{\Gamma(\kappa-3/2)}\left(1+\frac{v^2}{\kappa v_{te}^2}\right)^{-\kappa}
\end{equation}

Based on this, Equation~\ref{equ: nVF-ra} becomes:

\begin{equation}\label{equ: kappa-nVF}
\langle nVF(E) \rangle = EM \cdot \frac{2^{3/2}}{\sqrt{\pi m_e}} \cdot \frac{\Gamma(\kappa)}{\Gamma(\kappa-1.5)} \cdot \frac{\left(E/\kappa k_B T_\kappa\right)^{1.5}}{\left(1+E/\kappa k_B T_\kappa\right)^{\kappa}}
\end{equation}

The emission measure is calculated as $EM = n^2 V$, where $V = V_{\rm LT}$. The corresponding number density $n_\kappa$ is derived from:

\begin{equation}\label{equ: kappa-nk}
n_k = \frac{\dot{N}_0 (\kappa - 2) B(\kappa - 3/2, 1/2)}{2A \kappa^{1/2} \sqrt{\frac{2k_B T_\kappa}{m_e}}}
\end{equation}

where we use $A = A_{\rm LT}$ as the injection area.

The resulting DEM and mean electron flux spectrum from the WTM-derived kappa distribution are plotted as black curves in Figure~\ref{fig: spec_comb_0224}. The temperature and energy range below 6~keV—the lower limit of RHESSI fitting—are indicated with dashed lines to reflect limited observational constraints.

As seen in Figure~\ref{fig: spec_comb_0224}, the accelerated electron population inferred from the WTM is more than an order of magnitude smaller than the ambient thermal electron population derived from AIA. When comparing the two kappa distributions, the forward-fitted hot component has a lower temperature (1.06~keV vs. 1.29~keV for the WTM) and a significantly larger total emission measure ($5.9 \times 10^{47}$ vs. $1.8 \times 10^{45}$~cm$^{-3}$) than the WTM-derived accelerated electron spectrum. The emission measure fraction of the WTM-derived population relative to the forward-fit hot component is 0.3\%, corresponding to a number density ratio of approximately 5.5\%, assuming an identical source volume. This ratio offers a meaningful estimate of the fraction of flare-associated electrons that undergo acceleration. The emission measure and density derived from the WTM fit represent the energetic electron population, while those from the forward-fit DEM characterize the thermal plasma component associated with the same flaring region. Both diagnostics are applied consistently to the same source volume, so the emission measure ratio can be regarded as at least a conservative lower limit, providing a physically reliable proxy for the fraction of electrons participating in the acceleration process. Here we also note that the ratio between the thermal and nonthermal components obtained from the obtained kappa distribution can provide valuable information about the balance between acceleration and collisions, but it does not represent the fraction of the accelerated population and is therefore not the focus of this study.

\subsection{Physical Implications} \label{sec:20110224 anal}

As discussed above, the electron distributions obtained from WTM and EUV diagnostics provide valuable constraints on the physical conditions and processes occurring at the flare site. One key property evaluated in Section~\ref{sec:20110224 compare} is the fraction of accelerated electrons relative to the surrounding thermal population. In our earlier estimate, this ratio was calculated based on the injection area inferred from the looptop source ($A = A_{\rm LT}$) in determining the accelerated electrons (Equation \ref{equ: kappa-nk}). While the precise geometry of the acceleration region remains uncertain, the size of the footpoint source—typically rooted in the chromosphere, where the magnetic field strength is highest—can be used as a lower bound for the injection area ($A = A_{\rm FP}$). This suggests that the actual fraction of accelerated electrons could be as high as $\sim$7.0\%. Unless the acceleration occurs within a highly compact region—such as a cusp structure above the looptop \citep{2014ApJ...780..107K,2020NatAs...4.1140C,2024ApJ...971...85C,2021ApJ...911....4L}—with a much smaller injection area and correspondingly higher density, the accelerated electron fraction is unlikely to be significantly larger. For this event, no observational evidence suggests the existence of such a compact region. Future high-sensitivity HXR instruments may provide the resolution and dynamic range needed to directly detect such compact acceleration sites.

\begin{figure*}[!ht]
\plotone{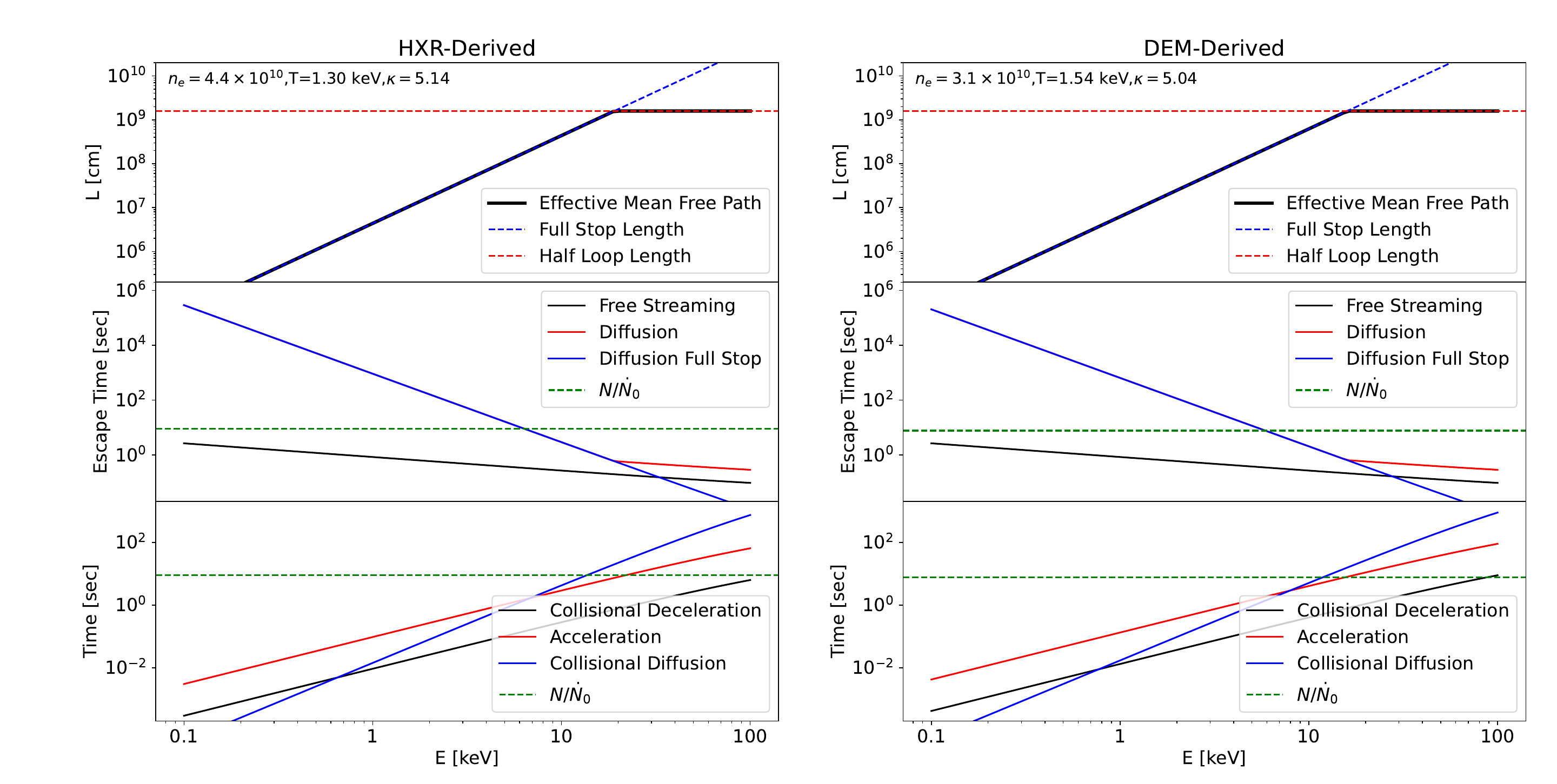}
\caption{Characteristic spatial scales and timescales for electron transport and acceleration during the 2011 February 24 flare. 
Left and right columns correspond to plasma properties derived from HXR spectral fitting and DEM analysis, respectively. 
Top panels show the electron mean free path $\lambda$ as a function of energy. The black curves denote the effective mean free path constrained to be no larger than the half loop length $L_{\rm loop}$, while blue dashed lines represent the collisional stopping length ($\lambda_{\rm stop} = E^2 / 2Kn$) as the full stop length.
Middle panels compare the observationally inferred escape time ($t_{\rm esp,obs} = N / \dot{N}_0$, green dashed), the free-streaming escape time ($L / v$, black), and diffusive escape time ($3L^{2} / \lambda v$) using either $\lambda_{\rm stop}$ (blue) or the constrained value (red).
Bottom panels show the collisional deceleration time $\tau_c$, stochastic acceleration time $\tau_{\rm acc} = 2\kappa \tau_c$, and collisional diffusion time $\tau_d = 2v^5 / \Gamma v_{te}^2$.}
\label{fig: time_0224}
\end{figure*}

In addition, characteristic timescales associated with electron acceleration and transport are of particular interest. We first estimate the total number of electrons in the source volume, inferred from DEM analysis as $N = n_e V = 2.01 \times 10^{37}$ electrons. The escape time of accelerated electrons can be approximated as $t_{\rm esp,obs} = N / \dot{N}_0$, where $\dot{N}_0$ is the injection rate from WTM spectral fitting. Using $\dot{N}_0 = 22.5 \times 10^{35}$~electrons~s$^{-1}$ (from the WTM fit using HXR-derived thermal properties), the escape time is 8.95~s. For the case using DEM-derived thermal properties, $\dot{N}_0 = 26.1 \times 10^{35}$~electrons~s$^{-1}$, yielding $t_{\rm esp,obs} = 7.72$~s.

For comparison, we also compute the energy-dependent free-streaming escape time $t_{\rm esp,free} = L / v(E)$, where $L = L_{\rm loop} = 15.8$~Mm and $v(E)$ is the electron speed at different energies. These results are plotted as black curves in the middle panels of Figure~\ref{fig: time_0224}. Moreover, under diffusive transport conditions, the escape time is approximated as $\tau_{\rm esp,diff} = 3L^{2} / \lambda v(E)$, where $\lambda$ is the electron mean free path. We adopt the mean free path $\lambda \sim \lambda_{\rm stop} = E^2 / 2Kn$ based on the collisional stopping length, shown as blue dashed lines in the top panels of Figure~\ref{fig: time_0224}. The left and right panels correspond to plasma properties from HXR and DEM diagnostics, respectively. We further constrain the mean free path such that $\lambda \leq L_{\rm loop}$, and the effective value is shown by the black lines in the top panels. The resulting diffusive escape times are shown as red curves in the middle panels, while the limiting case using the full stopping length as $\lambda$ is shown in blue. For reference, the observationally inferred escape time $t_{\rm esp,obs}$ is overplotted as a green dashed line. Figure~\ref{fig: time_0224} suggests that the calculated diffusive escape timescales are found to be in closer agreement with the observed $\frac{N}{\dot{N}_0}$ values for low-energy accelerated electrons (5-20~keV). Those electrons are more likely to propagate diffusively through the coronal loop, rather than freely streaming along magnetic field lines. These electrons interact more strongly with the ambient plasma, and their smaller velocities make them more susceptible to pitch-angle scattering. Consequently, they tend to remain confined longer and undergo thermalization. These results reinforcing the idea that these electrons are thermalized within the coronal plasma before escaping—consistent with the assumptions of the warm-target model.

To further assess the transport regime, we compute timescales for various physical processes: collisional deceleration $\tau_c \approx v^3 / \Gamma$, where $\Gamma = 2Kn / m_e^2$; stochastic acceleration $\tau_{\rm acc} = 2\kappa \tau_c$; and collisional diffusion $\tau_d = 2v^5 / \Gamma v_{te}^2$. These timescales are shown in the bottom panels of Figure~\ref{fig: time_0224}. Notably, the acceleration timescale is close to $t_{\rm esp,obs}$ in the 10–20~keV energy range, which roughly corresponds to the threshold above which electrons can escape into the dense chromosphere.

\section{2012 July 19 flare} \label{sec:20120719}

\subsection{Event overview and DEM}\label{sec:20120719 dem}
\begin{figure*}[!ht]
\plotone{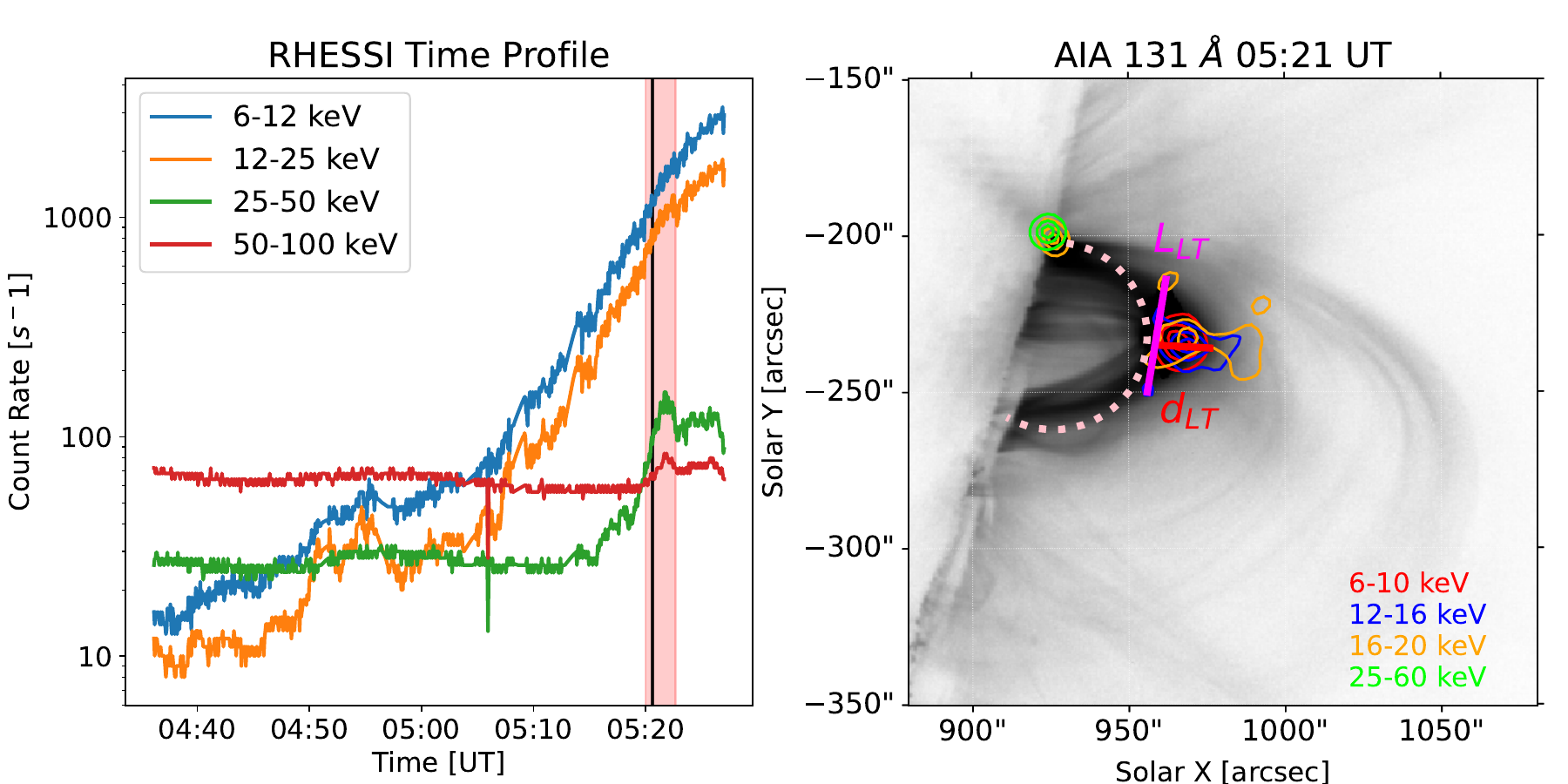}
\caption{Overview of the 2012 July 19 M7.7-class limb flare. 
Left panel: RHESSI light curves across different energy bands. The red shaded region indicates the time interval used for HXR spectral analysis, and the black shaded region marks the time of EUV DEM analysis.
Right panel: RHESSI CLEAN images in the 6–10~keV (red), 12–16~keV (blue), 16–20~keV (orange), and 25–60~keV (green) energy bands, overlaid on AIA 131~\AA\ context, at the 40\%, 70\%, and 90\% levels. The geometric parameters $d_{\rm{LT}}$ and $L_{\rm{LT}}$ are marked with red and magenta lines, respectively.}
\label{fig: event_0719}
\end{figure*}

\begin{figure*}[!ht]
\plotone{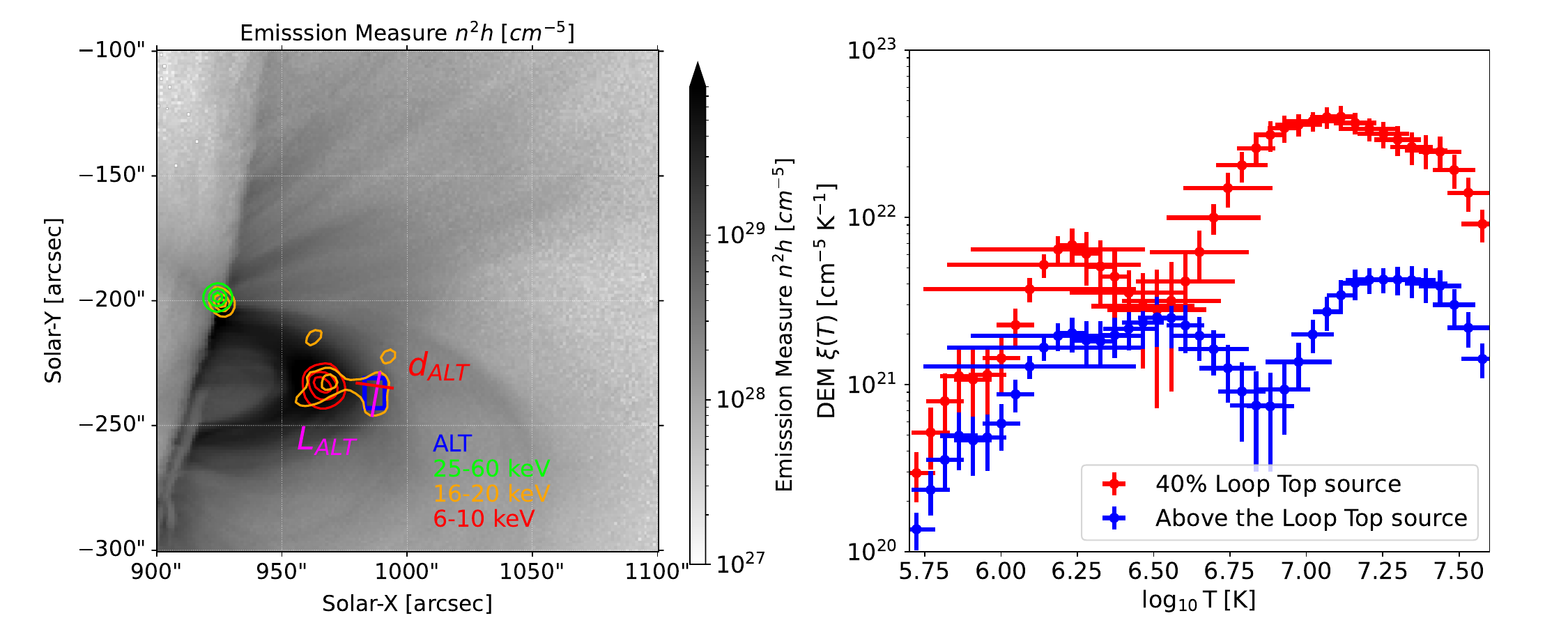}
\caption{DEM analysis for the 2012 July 19 flare. 
Left panel: Emission measure ($n^2h [\text{cm}^{-5}]$) map integrated from the regularized inversion. The geometric parameters $d_{\rm{ALT}}$, and $L_{\rm{ALT}}$ are marked with red and magenta lines, respectively.
Right panel: Resulting DEM distributions from the two regions. The red curve corresponds to the looptop source (6-10 keV 40\% red contour in the left panel), and the blue curve represents the ALT source (blue box in the left). }
\label{fig: dem_0719}
\end{figure*}

The other flare investigated in this study is a GOES M7.7-class limb event that occurred on 2012 July 19. This flare has been the subject of multiple studies \citep{2013ApJ...767..168L,2014ApJ...793L..23S,2019ApJ...872..204B}. The GOES SXR flux peaked at approximately 05:58~UT, while RHESSI was in orbital night from 05:23 to 06:11~UT. Consequently, we focus on the HXR burst occurring between 05:20:00 and 05:22:40~UT, which serves as the fitting interval for our spectral analysis (see the red shaded region in Figure~\ref{fig: event_0719}, left panel). RHESSI spectroscopic images across multiple energy bands are presented in Figure~\ref{fig: event_0719} (right panel); see also \citet{2014ApJ...780..107K} for more detailed HXR imaging analysis.

At the time of the selected burst, the 6--10~keV image (red contours) shows a classic looptop source, while the 12--16~keV (blue contours) and 16--20~keV (orange contours) images exhibit HXR emission distributed along the coronal loop, delineated by the 40\% contours. Similar to the 2011 February 24 event, we define the geometric parameters listed in Table \ref{tab: thermal_params_20120719}. The 25--60~keV image reveals a distinct northern footpoint source (green contours). \citet{2014ApJ...780..107K} reported the presence of a weak southern footpoint and an above-the-looptop (ALT) source using a two-step CLEAN algorithm. The ALT source is spatially consistent with the 16--18~keV secondary source (see their Figure~2). As we do not perform spatially resolved HXR spectral fitting in this work, we refer to the 16--20~keV secondary source located above the main looptop region as the ALT source. This region is likely located near the cusp structure of the flare and likely be closely associated with electron acceleration. The geometric parameters of the ALT source are also shown in Table \ref{tab: thermal_params_20120719}.

We conducted DEM analysis for both the looptop source (outlined by red contours in Figure~\ref{fig: dem_0719}, left panel) and the ALT source (enclosed by the blue box). The analysis was performed at 05:20:35~UT, coinciding with the HXR burst interval (black shaded region in Figure~\ref{fig: event_0719}, left panel). We used observations from six AIA EUV channels—94, 131, 171, 193, 211, and 335~\AA. The resulting DEM distributions are shown in Figure~\ref{fig: dem_0719} (right panel), with the looptop and ALT regions plotted in red and blue, respectively. All relevant parameters are also shown in Table \ref{tab: thermal_params_20120719}.

\begin{table}
    \centering
    \renewcommand{\arraystretch}{1.3}
    \begin{tabular}{||c|c|p{6.5cm}||}
        \hline
        \textbf{Parameter} & \textbf{Value} & \textbf{Remark} \\
        \hline\hline
        \multicolumn{3}{||c||}{\textit{Observation size}} \\
        \hline
        Looptop source diameter $d_{\rm LT}$ & $17.0\arcsec \ (= 12.6~\mathrm{Mm})$ & Defines column depth for AIA-DEM for the looptop region. \\
        Looptop source length $L_{\rm LT}$ & $36.5\arcsec \ (= 26.9~\mathrm{Mm})$ & Used to compute looptop source volume. \\
        \shortstack[c]{Looptop cross-sectional area\\ $A_{\rm LT} = \pi (d_{\rm LT}/2)^2$} & $1.2 \times 10^{18}$ cm$^2$ & Defines electron injection area at the looptop. \\
        \shortstack[c]{Looptop source volume\\ $V_{\rm LT} = \pi (d_{\rm LT}/2)^2 L_{\rm LT}$} & $3.3 \times 10^{27}$ cm$^3$ & Used for density and emission measure calculation. \\
        ALT source diameter $d_{\rm ALT}$ & $14.1\arcsec \ (= 10.4~\mathrm{Mm})$ & Defines column depth for AIA-DEM in the ALT region. \\
        ALT source length $L_{\rm ALT}$ & $17.3\arcsec \ (= 12.7~\mathrm{Mm})$ & Used to compute ALT source volume. \\
        \shortstack[c]{ALT source volume\\ $V_{\rm ALT} = \pi (d_{\rm ALT}/2)^2 L_{\rm ALT}$} & $1.1 \times 10^{27}$ cm$^3$ & Used for density and emission measure calculation in the ALT region. \\
        \hline

        \multicolumn{3}{||c||}{\textit{AIA-DEM-derived thermal properties}} \\
        \hline
        Looptop emission measure per area $EM_A$ & $8.7 \times 10^{29}$ cm$^{-5}$ & Used with $d_{\rm LT}$ to derive $n_e$. \\
        Looptop average temperature $T_{\rm avg}$ & $18.6$ MK ($=1.61$ keV) & Input for WTM loop thermal parameters. \\
        \shortstack[c]{Looptop electron number density\\ $n_e = \sqrt{EM_A / d}$ (with $d=d_{\rm LT}$)} & $2.6 \times 10^{10}$ cm$^{-3}$ & Input for WTM loop thermal parameters. \\
        ALT emission measure per area $EM_A$ & $1.1 \times 10^{29}$ cm$^{-5}$ & DEM for ALT region. \\
        ALT average temperature $T_{\rm avg}$ & $20.5$ MK ($=1.77$ keV) & Thermal properties for ALT region. \\
        ALT electron number density $n_e$ & $1.0 \times 10^{10}$ cm$^{-3}$ & Thermal properties for ALT region. \\
        Plausible temperature range 
        & $1.53$--$1.74$ keV (17.8--20.2 MK) 
        & Used to assess uncertainty of WTM spectral fits. \\
        Plausible density range 
        & $(2.2$--$3.4)\times 10^{10}$ cm$^{-3}$ 
        & Used to assess uncertainty of WTM spectral fits. \\
        \hline

        \multicolumn{3}{||c||}{\textit{HXR-derived thermal properties and plausible ranges}} \\
        \hline
        Isothermal component temperature
        & $2.04$ keV (23.6 MK) 
        & Used in WTM fit. \\
        Isothermal component electron density 
        & $1.7 \times 10^{10}$ cm$^{-3}$ 
        & Used in WTM fit. \\
        Plausible density range
        & ($1.3$--$2.0) \times 10^{10}$ cm$^{-3}$ 
        & Used to assess uncertainty of WTM spectral fits. \\
        Plausible temperature range 
        & $1.97$--$2.11$ keV (22.9--24.5 MK) 
        & Used to assess uncertainty of WTM spectral fits. \\
        \hline
    \end{tabular}
    \caption{Observation size, AIA-DEM-derived, and HXR-derived thermal parameters for the 2012 July 19 flare, with indication of their usage in the analysis.}
    \label{tab: thermal_params_20120719}
\end{table}

\subsection{HXR spectral analysis} \label{sec:20170719 hxr}

\begin{figure*}[!ht]
\plotone{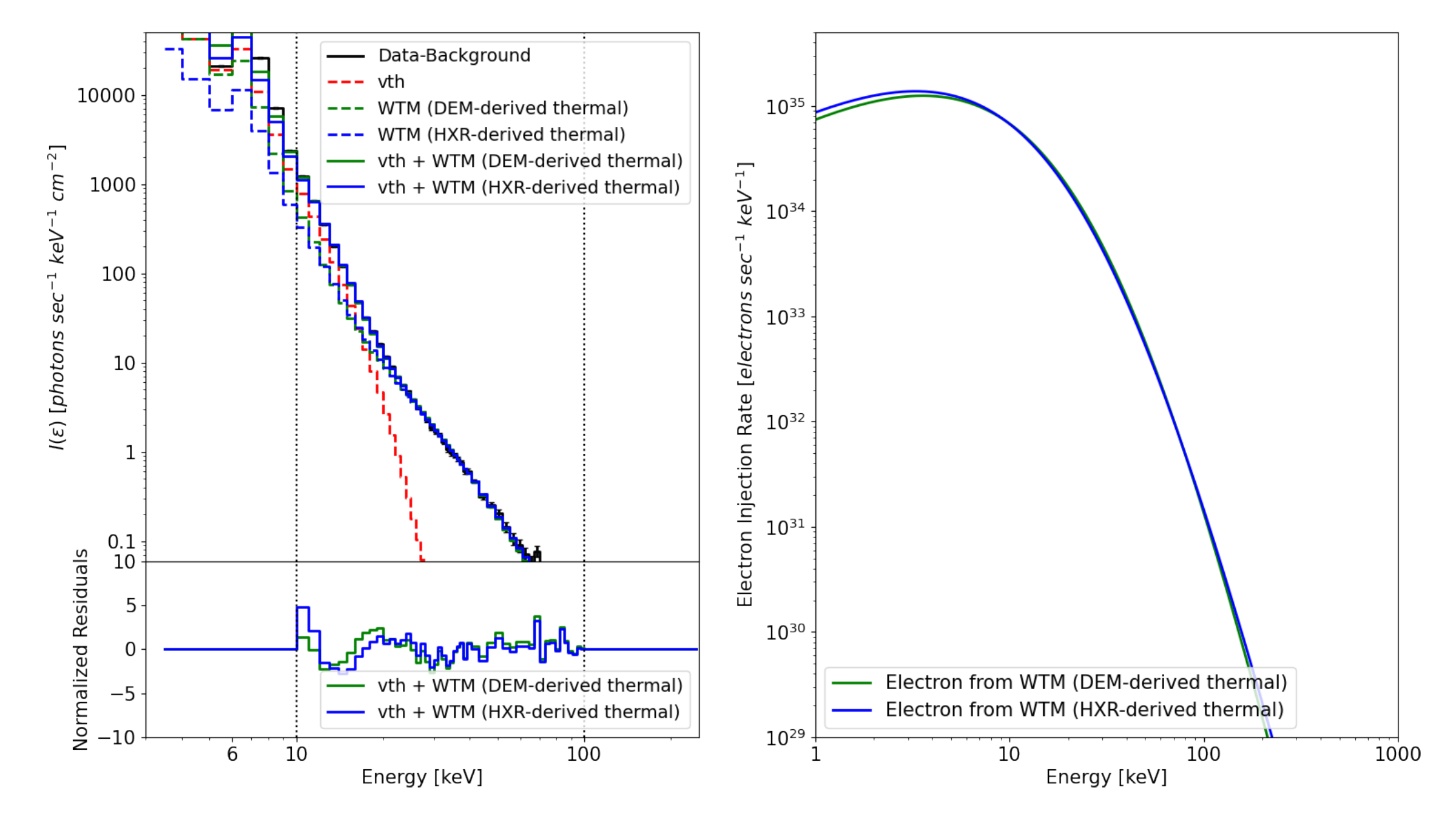}
\caption{Warm-target model fitting results for the 2012 July 19 M7.7-class flare.
Left panel: RHESSI photon spectrum fitted with the combined \textit{f\_vth} and \textit{f\_thick\_warm\_kappa} components. Blue curves correspond to the WTM fit results with thermal properties derived from the HXR isothermal fit, while green curves show the fit results use thermal input from the DEM analysis. Normalized residuals are shown in the lower sub-panel.
Right panel: Injected electron spectra corresponding to the two fits.}
\label{fig: wtm_res_0719}
\end{figure*}

\begin{table}
    \centering
    \begin{tabular}{||c|c|c|c||}
        \hline
        \textbf{Parameter} 
        & \shortstack[c]{\rule{0pt}{2.6ex}\textbf{$f_{\rm vth}$+WTM}\\\textbf{(thermal properties derived from HXR)}} 
        & \shortstack[c]{\rule{0pt}{2.6ex}\textbf{$f_{\rm vth}$+WTM}\\\textbf{(thermal properties derived from DEM)}} 
        & \textbf{Status} \\
        \hline\hline
        \multicolumn{4}{||c||}{\textit{\boldmath $f_{\rm vth}$ (isothermal component, fixed)}} \\
        \hline
        EM [$10^{49}\,\mathrm{cm}^{-3}$]   & 0.098  & 0.098  & Fixed \\
        $k_{B}T$ [keV]                     & 2.04   & 2.04   & Fixed \\
        \hline
        \multicolumn{4}{||c||}{\textit{WTM component: loop thermal parameters (fixed)}} \\
        \hline
        $n_{\rm loop}$ [$10^{10}\,\mathrm{cm}^{-3}$] & 1.7  & 2.6  & Fixed \\
        $k_{B}T_{\rm loop}$ [keV]                    & 2.04 & 1.61 & Fixed \\
        $L$ [Mm]                                     & 38.9 & 38.9 & Fixed \\
        \hline
        \multicolumn{4}{||c||}{\textit{WTM component: kappa parameters related to accelerated/injected electrons (free)}} \\
        \hline
        $\dot{N}_0$ [$10^{35}\,\mathrm{e^-}\,\mathrm{s}^{-1}$] & $15.3 \pm 0.6$ & $14.7 \pm 0.5$ & Free \\
        $\kappa$                                                & $8.31 \pm 0.19$ & $8.97 \pm 0.21$ & Free \\
        $k_{B}T_{\kappa}$ [keV]                                 & $2.89 \pm 0.09$ & $3.17 \pm 0.09$ & Free \\
        \hline
        \multicolumn{4}{||c||}{\textit{Fit diagnostics}} \\
        \hline
        $\chi^2$      & 2.04 & 1.89 & N/A \\
        Power [$10^{28}\,\mathrm{erg\,s}^{-1}$] & 2.20 & 2.24 & N/A \\
        \hline
    \end{tabular}
    \caption{Same as in Table~\ref{tab: 20110224 fit} for the 2011 February 24 flare, spectral fitting results for the 2012 July 19 M7.7-class flare using the warm-target model with a kappa-form injected electron distribution and an isothermal component ($f_{\rm vth}$). Two fits are shown that differ in the thermal properties specified for the WTM: one set derived from an HXR isothermal fit, and the other from DEM analysis of the loop-top region. Fixed parameters, free parameters, and diagnostic quantities are indicated.}
    \label{tab: 20120719 fit}
\end{table}

We note that this burst occurs before the impulsive phase and is characterized by relatively weak coronal HXR sources together with chromospheric footpoint HXR emission, differing from the typical flare geometry. We propose that these are possible thick-target coronal HXR sources \citep[e.g.,][]{2008A&ARv..16..155K,2016ApJ...816...62F,2018ApJ...867...82D}, which likely mark the region of electron acceleration. For this burst, the accelerated electrons still escape from the corona and propagate downward to the chromosphere, while a fraction of them thermalize within the corona, as supported by the presence of 16–20~keV sources along the coronal loops. Therefore, we suggest that the application of the WTM in spectral analysis remain valid for this event.

Same as in the analysis of the 2011 February 24 flare, we perform HXR spectral fitting for this event using the warm-target model with a kappa-form injected electron distribution (\textit{f\_thick\_warm\_kappa}), combined with an isothermal thermal component (\textit{f\_vth}). The thermal properties of the target coronal loop are derived through two independent approaches. The half-loop length $L_{\rm loop}$ is estimated to be 38.9~Mm, as indicated by the pink dashed line in Figure~\ref{fig: event_0719} (right panel). From the fitted isothermal component during the pre-burst interval (05:19:52–05:20:00~UT), we obtain an emission measure of $0.098 \times 10^{49}$~cm$^{-3}$. Assuming a source volume $V = V_{\rm LT} = 3.3 \times 10^{27}$~cm$^3$, this corresponds to an electron number density of $n_{\rm loop} = 1.7 \times 10^{10}$~cm$^{-3}$, with a temperature of $k_{\rm B} T_{\rm loop} = 2.04$~keV. Meanwhile, the DEM analysis for the looptop region from AIA observations yields $n_e = 2.6 \times 10^{10}$~cm$^{-3}$ and $k_{\rm B} T_{\rm loop} = 1.61$~keV. The resulting WTM fit parameters are summarized in Table~\ref{tab: 20120719 fit}, and the corresponding photon and electron spectra are shown in Figure~\ref{fig: wtm_res_0719}. The fits from both sets of thermal inputs show good consistency.

Following the same approach as for the 2011 event, we define a plausible range for the thermal parameters by combining the results from the pre-burst isothermal fit and AIA DEM analysis at nearby times. From the isothermal component fitted during 05:19:52--05:20:00 UT, we obtain a temperature range of 1.97--2.11 keV and an emission measure of $(0.057$--$0.139)\times10^{49}$ cm$^{-3}$, corresponding to an electron density of $(1.3$--$2.0)\times10^{10}$ cm$^{-3}$. From the DEM analysis at 05:18:35 UT and 05:21:47 UT for the looptop region, we obtain a temperature range of 1.53--1.74 keV and density range of $(2.2$--$3.4)\times10^{10}$ cm$^{-3}$. Combining these, we adopt a thermal input range of 1.53--2.11 keV in temperature and $1.3\times10^{10}$--$3.4\times10^{10}$ cm$^{-3}$ in electron density. Within these ranges, the mean values of the fitted kappa parameters are: $\dot{N}_0 = 15.3 \times 10^{35}$ electrons~s$^{-1}$, $k_{\rm B} T_{\kappa} = 3.35$~keV, and $\kappa = 9.11$. The corresponding standard deviations are $9.3 \times 10^{35}$~electrons~s$^{-1}$, 0.80~keV, and 1.04, respectively. As in the earlier event, the injection rate $\dot{N}_0$ is relatively less constrained primarily influenced by a few extreme large values. Nevertheless, the warm-target model reliably constrains the overall shape of the kappa-form electron distribution across the examined thermal parameter space.

\subsection{Comparison of Electron Spectra from WTM and EUV Diagnostics} \label{sec:20120719 compare}

\begin{figure*}[!ht]
\plotone{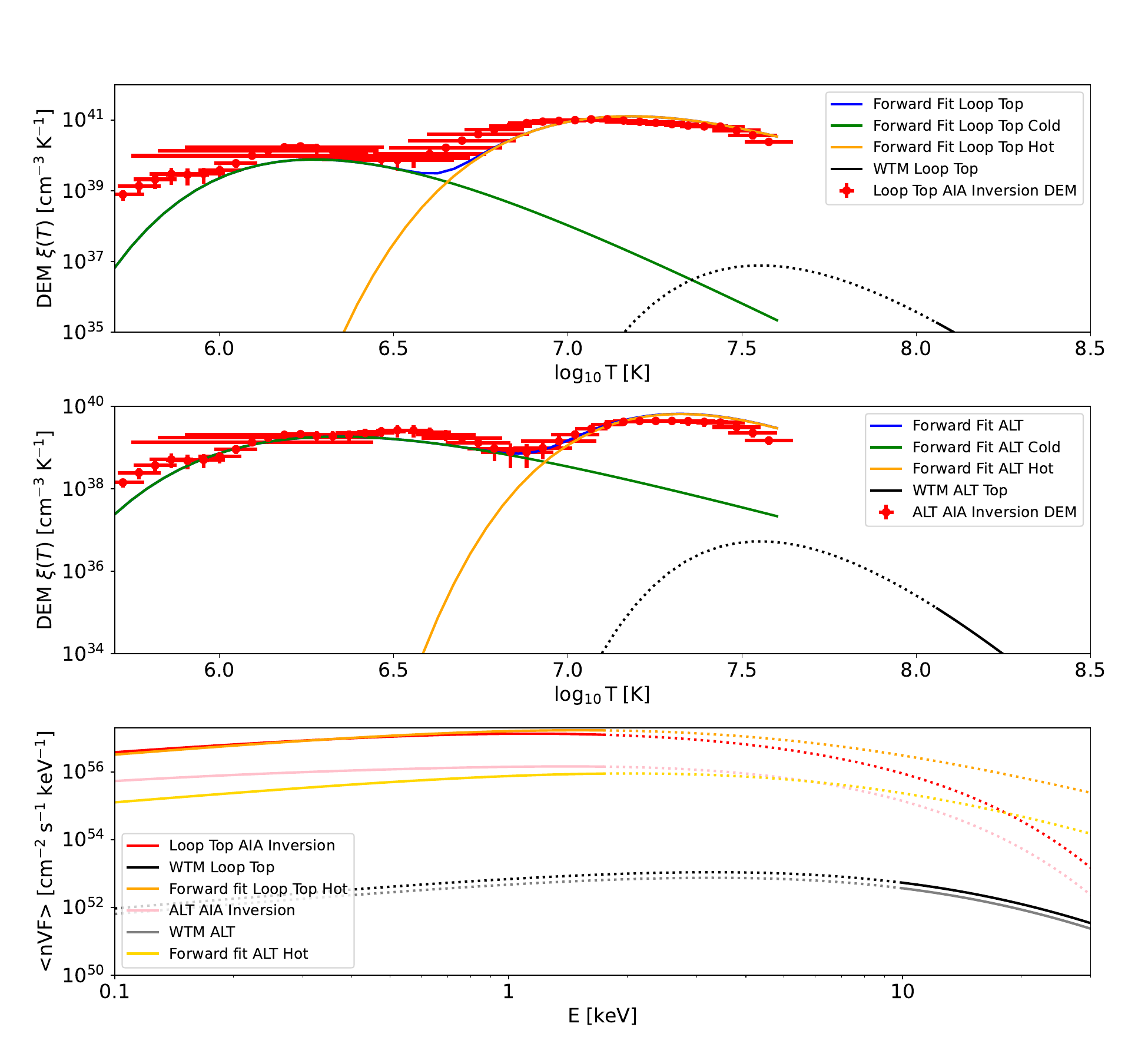}
\caption{Top: Comparison of DEM distributions derived from the AIA looptop region using regularized inversion (red) and forward fitting (blue), alongside the DEM inferred from the WTM kappa-form electron spectrum (black). For the WTM-derived DEM, values below 10~keV—the lower limit of the spectral fit—are indicated with a dashed line. 
Middle: Same as the top panel, but for the above-the-looptop (ALT) region.
Bottom: Corresponding mean electron flux spectra $\langle nVF \rangle(E)$ derived from the AIA DEMs (regularized inversion and forward fit) and from the WTM-based kappa distribution. For the AIA-derived spectra, energies corresponding to temperatures above 20~MK (beyond AIA’s sensitivity) are shown as dashed extensions. The WTM-derived electron flux below 10~keV is also represented with a dashed line. Results from both the looptop and ALT regions are shown.}
\label{fig: spec_comb_0719}
\end{figure*}

The same forward-fitting procedure, using two kappa-form DEM components, is applied to the AIA observations for this flare. The fitting is performed separately for the looptop region and the above-the-looptop (ALT) region. The fitted DEM distributions and corresponding mean electron flux spectra are shown in Figure~\ref{fig: spec_comb_0719}.
For the looptop source, the fitted parameters are: 
$\text{EM}_\text{cold} = 2.1 \times 10^{46}$~cm$^{-3}$, 
$k_{\rm B} T_{\kappa\text{-cold}} = 0.15$~keV (fixed), 
$\kappa_{\text{cold}} = 5.5$, 
$\text{EM}_\text{hot} = 3.5 \times 10^{47}$~cm$^{-3}$, 
$k_{\rm B} T_{\kappa\text{-hot}} = 1.14$~keV, 
and $\kappa_{\text{hot}} = 4.3$.
For the ALT region, the fitted parameters are: 
$\text{EM}_\text{cold} = 1.2 \times 10^{46}$~cm$^{-3}$, 
$k_{\rm B} T_{\kappa\text{-cold}} = 0.16$~keV (fixed), 
$\kappa_{\text{cold}} = 2.8$, 
$\text{EM}_\text{hot} = 2.0 \times 10^{47}$~cm$^{-3}$, 
$k_{\rm B} T_{\kappa\text{-hot}} = 1.65$~keV, 
and $\kappa_{\text{hot}} = 5.3$.

We note that \citet{2014ApJ...780..107K} also modeled the DEM distribution as a combination of cold and hot components to reproduce AIA counts in the ALT region. While our study uses a different DEM functional form and assumes different source volumes, leading to discrepancies in the total emission measures, the peak temperatures of both cold and hot components are broadly consistent with their results—indicating general agreement in the derived thermal structure. Different from the previous study by \citet{2013ApJ...779..107B}, which analyzed the RHESSI spectrum and AIA observations in the looptop and ALT regions to assess the instantaneous local electron distribution, our approach focuses on the flare-integrated spectrum. The decision to utilize an integrated spectrum arises from the inherent uncertainties associated with RHESSI imaging spectroscopy, where the CLEAN algorithm can introduce uncertainties of up to $\sim$10\% \citep{2002SoPh..210...61H,2004SoPh..219..149A}. Instead, we determine the flare-accelerated electrons more precisely using the integrated HXR spectrum. For this event, we suggest that the ALT region is likely the primary site of electron acceleration, whereas the loop-top region represents the injection site of the accelerated electrons. According to our fits, the ratio of the accelerated electron number density (derived from the WTM fit) to the ambient thermal electron population (from the hot kappa component) is approximately 0.9\% in the looptop region and 3.1\% in the ALT region. The higher ratio in the ALT region suggests a greater acceleration efficiency or a larger fraction of electrons being energized, assuming that the ALT region, rather than the loop-top region, is the primary acceleration site. This interpretation is consistent with previous studies of particle acceleration in ALT regions \citep[e.g.,][]{2010ApJ...714.1108K,2020NatAs...4.1140C}. We expect that future spatially resolved observations will allow a more detailed investigation of the acceleration rate and its location.

\subsection{Physical Implications} \label{sec:20120719 phys}

\begin{figure*}[!ht]
\plotone{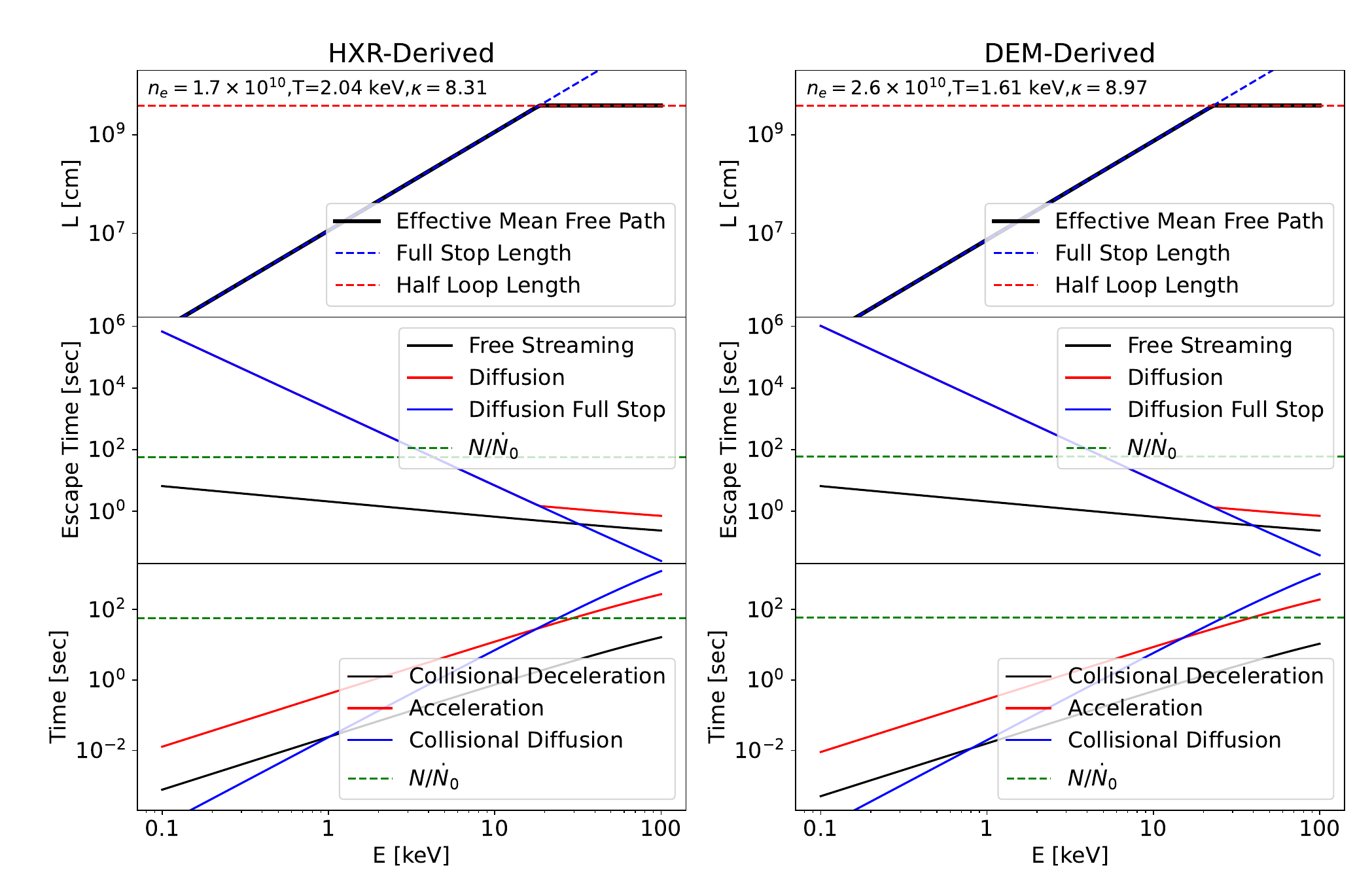}
\caption{Characteristic spatial scales and timescales for electron transport and acceleration for the 2012 July 19 flare. 
Top panels: effective mean free path (black curves) used in the collisional diffusion calculation, based on the stopping length limited to $L_{\rm loop}$ (blue dashed lines). Left and right panels correspond to thermal properties derived from HXR and DEM diagnostics, respectively.
Middle panels: energy-dependent escape times under free-streaming (black) and collisional diffusion transport (red for $\lambda = \min(\lambda_{\rm stop}, L_{\rm loop})$, blue for $\lambda = \lambda_{\rm stop}$). Escape times inferred from $\frac{N}{\dot{N}_0}$ are marked by green dashed lines.
Bottom panels: timescales for collisional deceleration ($\tau_c$), stochastic acceleration ($\tau_{\rm acc}$), and collisional diffusion ($\tau_d$), calculated under both thermal conditions.}
\label{fig: time_0719}
\end{figure*}

Same as for the 2011 February 24 flare, we also examine the characteristic timescales associated with electron acceleration and transport for the 2012 July 19 event. The total number of electrons within the source volume is estimated to be $N = n_e V = 8.78 \times 10^{37}$ electrons. This yields escape times of 57.4 and 59.7 seconds, calculated using $N / \dot{N}_0$ from WTM fits with thermal properties derived from HXR and DEM diagnostics, respectively. Figure~\ref{fig: time_0719} presents the energy-dependent escape times for both free-streaming and collisional diffusion regimes. The results are consistent with those obtained for the 2011 flare, reinforcing the interpretation that electron transport occurs primarily through diffusion within the coronal loop. Additionally, the acceleration timescale is found to be comparable to $N / \dot{N}_0$ at around 30~keV， close to the critical escape energy  $E = \sqrt{2KnL_{\rm loop}} \approx 23.1$~keV.

\section{Conclusion and Discussion} \label{sec:conclusion}  

In this study, we investigated two well-observed GOES M-class limb flares using RHESSI hard X-ray (HXR) spectroscopy and SDO/AIA EUV observations. For the HXR spectral analysis, we applied the warm-target model (WTM) with kappa-distributed injected electrons to characterize the accelerated nonthermal electron population. In parallel, we performed differential emission measure (DEM) analysis using both regularized inversion and forward-fitting algorithms to study the thermal plasma associated with the flare, spanning temperatures from below 1~MK to $\sim$20~MK. Our results demonstrate that the combined WTM and DEM approaches yield robust constraints on the flare-associated energetic electrons, offering new insight into electron acceleration and transport.

As discussed, the motivation for adopting the WTM arises from its physically realistic representation of flare conditions and progress compared to the traditional cold-target model. By considering the thermalization of accelerated electrons in hot coronal loops, the WTM provides improved constraints on the nonthermal electron distribution, particularly in the $\lesssim$20~keV range where the cold-target approximation becomes unreliable. The effectiveness of the WTM in determining these electrons was thoroughly examined in \citealt{2019ApJ...871..225K,2024ApJ...974..119L}. Here, we further consider the impact of uncertainties in the thermal properties of the target coronal loops, as these influence the thermalization and can thus affect the fitted electron distributions. We perform the WTM fit within a physically plausible range of thermal parameters obtained by both HXR and EUV diagnostics. For the 2011 February 24 flare, we obtained best-fit parameters of $\dot{N}_0 = (25.9 \pm 12.5) \times 10^{35}$~electrons~s$^{-1}$, $k_{\rm B} T_\kappa = 1.27 \pm 0.25$~keV, and $\kappa = 5.11 \pm 0.06$. For the 2012 July 19 flare, the corresponding values were $\dot{N}_0 = (15.3 \pm 9.3) \times 10^{35}$~electrons~s$^{-1}$, $k_{\rm B} T_\kappa = 3.35 \pm 0.80$~keV, and $\kappa = 9.11 \pm 1.04$. These results reinforce the robustness of the WTM in reliably characterizing nonthermal electrons, even when the thermal environment is not precisely known.

The use of a kappa-form accelerated electron distribution in the WTM allows for meaningful direct comparisons with DEM results from AIA, including for energies below RHESSI’s detection threshold ($\sim$3~keV). For the AIA data, we employ both regularized inversion and forward-fitting techniques to obtain the DEM distribution. We propose that the hot kappa component in the forward-fit DEM more accurately captures the plasma heated during flares while effectively distinguishing it from cooler background plasma along the line of sight. The comparison between WTM-derived and AIA-derived electron distributions reveals that the number of flare-accelerated electrons constitutes only a small fraction of the ambient thermal population: $\sim$5.5\% for the 2011 February 24 flare, and $\sim$0.9\% and 3.1\% for the loop-top and above-the-loop-top regions in the 2012 July 19 event, respectively. These values are consistent with prior and recent studies \citep{2013ApJ...764....6O,2023ApJ...947L..13K,2025ApJ...987..211B,2025ApJ...990..101V}, supporting the algorithm in which nonthermal electrons, while energetically significant, make up a minor fraction of the total particle population in flaring regions.

We also investigated essential physical properties relevant to the processes of energy-containing electron acceleration and transport during flares. Comparing the inferred electron escape time ($N/\dot{N}_0$) with theoretical escape times under free-streaming and collisional diffusion, we find that low-energy electrons are more likely to propagate diffusively within the flare loops. This supports the physical framework of the warm-target model, which predicts thermalization of low-energy injected electrons in the corona. Additionally, we find that the characteristic acceleration timescale is comparable to $N/\dot{N}_0$ near 10–30~keV—close to the threshold energy above which electrons can escape into the dense chromosphere.

In this study, we focus exclusively on AIA observations to derive the thermal plasma properties at the flare site. While this provides weaker constraints on the high-temperature component, it provides essential additional constraint on the flaring plasma. 
We acknowledge that observations from other instruments, 
such as XRT \citep{2007SoPh..243...63G}, MaGIXS \citep{2019ApJ...884...24A,2022JAI....1150010C,2023ApJ...945..105S}, GOES, and the SXR component from HXR observations, can provide valuable additional insights. Previous studies \citep{2015Ge&Ae..55..995M,2014ApJ...789..116I,2019ApJ...872..204B} combining AIA and RHESSI to derive DEMs have shown some success; however, one must remain cautious about the sensitivity range discrepancies among instruments when performing such combined analyses. This raises the question of whether different diagnostics truly probe the same electron population 
or the spatially different as suggested by \citet{2015A&A...584A..89J}, 
complicating the use of an identical electron distribution in simultaneous forward fitting. 
The multi-instrument approach is particularly compelling, especially with the availability of more advanced tools, and we plan to incorporate such methods in future work to achieve better constraints.

It is important to note that the present study is based on spatially integrated HXR spectra. Future instruments with improved dynamic range and high-resolution imaging spectroscopy would be valuable for resolving the spatial structure of the acceleration region and assessing how injected electron populations evolve. Even so, our results highlight how combining WTM and DEM diagnostics enables detailed characterization of energy-containing electrons in flares. It is also valuable to consider how kappa-distributed injected electrons evolve as they inject into the coronal loops. Electrons at different energies experience distinct transport processes, and the resulting in-loop distribution may differ from the initially injected kappa form. Understanding this evolution provides a promising avenue for future studies aimed at revealing how flare-accelerated electrons redistribute and dissipate their energy in the coronal environment.

\begin{acknowledgments}
The work is supported via the STFC/UKRI grants ST/T000422/1 and ST/Y001834/1. We thank Galina Motorina for valuable discussions on the DEM analysis. We also thank the anonymous referee for constructive comments that improved the manuscript.
\end{acknowledgments}

\bibliography{sample7}{}
\bibliographystyle{aasjournalv7}

\end{CJK*}
\end{document}